\documentclass[twocolumn,showpacs,preprintnumbers,amsmath,amssymb,showkeys]{revtex4}
\usepackage{graphicx}
\usepackage{dcolumn}
\usepackage{bm}

\begin{document}

\title{Damping effects in hole-doped graphene: the relaxation-time approximation}
 \author{I. Kup\v{c}i\'{c}}     % Do not remove
 \address{
   Department of Physics, Faculty of Science, University of Zagreb, 
   P.O. Box 331,  HR-10002 Zagreb,  Croatia}

\begin{abstract} 
The dynamical conductivity of interacting multiband electronic systems derived in Ref.~\onlinecite{Kupcic13} is shown
to be consistent with the general form of the Ward identity.
	Using the semiphenomenological form of this conductivity formula, we have demonstrated that the relaxation-time 
approximation can be used to describe the damping effects in weakly interacting multiband systems only if
local charge conservation in the system
and gauge invariance of the response theory are properly treated.
	Such a gauge-invariant response theory is illustrated on the common tight-binding model for conduction 
electrons in hole-doped graphene.
	The model predicts two distinctly resolved maxima in the energy-loss-function spectra.
	The first one corresponds to the intraband plasmons (usually called the Dirac plasmons). 
	On the other hand, the second maximum ($\pi$ plasmon structure) is simply a consequence of
the van Hove singularity in the single-electron density of states.
	The dc resistivity and the real part of the dynamical conductivity are found to be well described 
by the relaxation-time approximation, but only in the parametric space in which the damping is dominated 
by the direct scattering processes.
	The ballistic transport and the damping of Dirac plasmons are thus the questions that
require abandoning the relaxation-time approximation.

\end{abstract}
%
% 78.67.Wj - Optical properties of graphene
% 72.80.Vp - Electronic transport of graphene
% 73.22.Pr - Electronic structure of graphene
% 71.45.Gm - [Collective effects] ..., plasmons
%
\maketitle

\section{Introduction}
Vertex corrections are the key to quantitative understanding of
both transport phenomena and low- and high-energy electron-hole and collective excitations in solids.
\cite{Abrikosov75,Mahan90}
	Their role becomes even more pronounced when the system under consideration has several bands at 
the Fermi level and in addition the electrical conductivity is low-dimensional. \cite{Dzyaloshinskii73,Solyom79}
	Therefore, graphene is ideally suited for studying
the effects associated with different types of vertex corrections.
	Graphene is a two-dimensional material with two $\pi$ bands in the vicinity of the Fermi level 
in which the (electron or hole) doping level can be easily tuned by the electric field effect.
\cite{Novoselov05,Zhang05,Li08}
	There is  a relatively good understanding of the single-electron properties based on the detailed 
angle-resolved photoemission spectroscopy (ARPES) investigations on pure, doped, and even heavily doped samples.
	The comparison of the single-electron Green's functions extracted from ARPES \cite{Bostwick07}
and the electron-hole propagators extracted from resistivity and reflectivity measurements \cite{Novoselov05,Zhang05,Li08}
as well as from electron-loss spectroscopy experiments \cite{Eberlein08,Yan13}
provides direct information about the nature of the electron-electron interactions and about the role of vertex corrections in different response functions.

From the theoretical standpoint, it is essential to use the response theory which treats the single-electron self-energy contributions and the vertex corrections on the same footing.
	If the relaxation processes in the system under consideration are related predominantly to the scattering 
by impurities, the standard method of impurity-averaged propagators can be applied. \cite{Abrikosov75,Vollhardt80,Rammer04}
	However, if the interactions (bare or renormalized) are retarded, we usually end up analyzing Bethe--Salpeter
equations (or the related quantum transport equations) in a way consistent with
the Dyson equations for electrons and phonons. \cite{Kupcic13,KupcicUP}

In graphene, conduction electrons are assumed to be weakly interacting, and, in principle, one can use the approximate solution 
of the Bethe--Salpeter equations in which the electron-hole self-energy is replaced by the memory function, or even by the frequency-independent relaxation rate. \cite{Castro09}
	In this paper, the quantum transport equations from Refs.~\onlinecite{Kupcic13} and \onlinecite{KupcicUP} 
are applied to hole-doped graphene.
	The dispersions of the electron-hole excitations and of the collective plasmon excitations are calculated 
beyond the Dirac cone approximation.
	For the purpose of comparison with the previous work, the damping effects are introduced in the semiphenomenological way.
	The vertex corrections are implicitly included
through the general Ward identity relations, which connect three types of the RPA (random phase approximation) 
irreducible real-time correlation functions.
	These relations are interesting in themselves because they take care of both
local charge conservation in the system and gauge invariance of the response theory.
	The detailed microscopic analysis of the intraband memory function in doped graphene, which 
is an obvious generalization of the intraband relaxation rate, will be given in the accompanying article. \cite{KupcicUP}

Precisely speaking, this paper is devoted to the electrodynamic properties of weakly interacting multiband 
electronic systems described by an exactly solvable bare Hamiltonian in the case
in which the Lorentz local field corrections are negligible.
	The Hamiltonian includes also the retarded phonon-mediated electron-electron interactions, 
the non-retarded long-range and short-range Coulomb interactions, the electron scattering processes 
from static disorder, and the coupling to external fields.
	We shall label the microscopic longitudinal dielectric function by $\varepsilon ({\bf q}, \omega)$, 
with the macroscopic dielectric function being its value at ${\bf q} \approx 0$.
	This function is given by \cite{Kubo95,Landau95,Kupcic13}
\begin{eqnarray}
&& \hspace{-13mm}
\varepsilon ({\bf q},\omega) \approx 1 - v({\bf q}) \chi({\bf q},\omega)
\nonumber \\
&& \hspace{1mm}
\approx \varepsilon_\infty ({\bf q}, \omega) - v({\bf q}) \chi^{\rm tot}({\bf q},\omega)
\nonumber \\
&& \hspace{1mm}
\approx \varepsilon_\infty ({\bf q}, \omega) + v({\bf q})\sum_{\alpha \beta} \frac{{\it i}}{\omega} 
q_\alpha \sigma_{\alpha \beta}^{\rm tot} ({\bf q},\omega) q_\beta,
\label{eq1}
\end{eqnarray}
where the dielectric susceptibility of interest
$\chi^{\rm tot}({\bf q}, \omega)=\chi^{\rm intra} ({\bf q}, \omega)+\chi^{\rm inter}({\bf q}, \omega)$
is the sum of the intraband and interband contributions,
and $\sigma_{\alpha \beta}^{\rm tot} ({\bf q}, \omega)$ is the corresponding conductivity tensor.
	Here, $\varepsilon_\infty ({\bf q}, \omega)$ describes both
the contributions originating from the rest of the high-frequency excitations and 
the local field corrections to $\sigma_{\alpha \beta}^{\rm tot} ({\bf q}, \omega)$.

For ${\bf q}$ not too large, the problem of calculating $\varepsilon ({\bf q}, \omega)$
in the gauge-invariant manner reduces to determining the gauge-invariant form of the conductivity tensor.
	Therefore, the general relations connecting the charge and current density fluctuations 
and the causality principle requirement are an essential part of a proper theoretical description
of both the low- and high-energy electrodynamic properties of such a system, including the damping of 
different types of elementary excitations.
	Pure and doped graphene are both very interesting two-band examples in which 
$\varepsilon_\infty ({\bf q}, \omega)$ can be approximated by the real constant $\varepsilon_\infty$, 
at least for $\hbar \omega < 5$ eV, and the total Hamiltonian includes, in principle, all aforementioned contributions. \cite{Castro09,Peres08}

In Sec.~II we consider the total Hamiltonian in graphene beyond the Dirac cone approximation 
and show all elements in it in the representation which is commonly used in the analysis of multiband electronic systems.
	In Secs.~III and IV the Ward identity relations are derived in the multiband case in which local field effects in 
$\sigma_{\alpha \beta} ({\bf q}, \omega)$ are negligible.
	In Secs.~V$-$VII the results are combined with the relaxation-time approximation 
to obtain the consistent description of the dynamical conductivity and the Dirac and $\pi$ plasmons in hole-doped graphene.
	Section VIII contains concluding remarks.

\section{Hole-doped graphene}
In hole-doped graphene conduction electrons are described by the Hamiltonian \cite{Castro09,Peres08}
\begin{eqnarray}
&& \hspace{-2mm}
H = H^{\rm el}_0 + H^{\rm ph}_0 + H'_1 + H'_2 + H^{\rm ext}.
\label{eq2}
\end{eqnarray} 
	$H$ is shown here in two representations commonly used in multiband electronic systems, 
in the diagonal Bloch representation $\{ L {\bf k} \}$ and in the representation 
of the delocalized orbitals $\{ l {\bf k} \}$. \cite{Kupcic13}
	For example, the bare electronic contribution $H^{\rm el}_0$, which represents an exactly solvable two-band 
tight-binding problem, takes the form
\begin{eqnarray}
&& \hspace{-10mm}
H_0^{\rm el} = \sum_{ll'} \sum_{{\bf k} \sigma} H^{ll'}_0 ({\bf k}) 
c^\dagger_{l {\bf k} \sigma} c_{l' {\bf k} \sigma}
=  \sum_{L{\bf k} \sigma} \varepsilon_L ({\bf k}) c^\dagger_{L {\bf k} \sigma} c_{L {\bf k} \sigma}.
\label{eq3}
\end{eqnarray}
	Here, $ c^\dagger_{l n \sigma}$ and
\begin{equation}
c^\dagger_{L {\bf k} \sigma} = \frac{1}{\sqrt{N}} \sum_{ln} e^{i {\bf k} \cdot {\bf R}_n} 
U_{\bf k} (L,l) c^\dagger_{l n \sigma} 
= \sum_l U_{\bf k} (L,l) c^\dagger_{l {\bf k} \sigma}
\label{eq4}
\end{equation}
are, respectively, the electron creation operators  in the $l$th orbital in the unit 
cell at the position ${\bf R}_n$ and in the band labeled by the band index $L$.
	The $U_{\bf k} (L,l)$ are the elements of the transformation matrix which connects these two representations.

The change to the $\{ s {\bf k} \}$ representation, which is widely used in the literature focused 
on the physics of graphene, is straightforward.
	The index $l = A, B$ labels two different $2p_z$ orbitals on two carbon sites in the unit cell, and 
the band index $s = \pi^*, \pi$ (or $s = +1,-1$) labels the corresponding $2p_z$ bands.
	The relevant matrix elements are $H^{ll}_0 ({\bf k}) = \varepsilon_{p_z}=0$ and
$H^{BA}_0 ({\bf k}) = t({\bf k})$, resulting in 
\begin{eqnarray}
&& \hspace{-10mm}
H_0^{\rm el} = \sum_{s = \pi^*, \pi} \sum_{{\bf k} \sigma} \varepsilon_s({\bf k}) c^\dagger_{s {\bf k} \sigma} 
c_{s {\bf k} \sigma},
\label{eq5}
\end{eqnarray}
where
\begin{eqnarray}
&& \hspace{-10mm}
\varepsilon_{\pi^*} ({\bf k}) =  |t ({\bf k})|, 
\hspace{10mm}
\varepsilon_{\pi} ({\bf k}) = - |t ({\bf k})|,
\label{eq6}
\end{eqnarray}
$t ({\bf k}) = -\sum_{j=1}^3 t_j \, e^{-i {\bf k} \cdot {\bf r}_j }$, and
\begin{equation}
|t ({\bf k})| = t \sqrt{ 3+ 2 \cos k_xa + 4 \cos \frac{k_x a}{2} \cos \frac{\sqrt{3} k_ya}{2} }.
\label{eq7}
\end{equation}
	The transformation matrix elements $U_{\bf k} (s,l)$ are given by Eq.~(\ref{eqC1}).

 \begin{figure}[tb]
  \includegraphics[width=15pc]{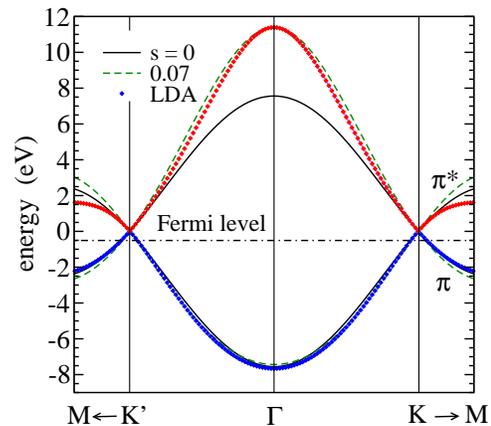}
  \caption{(Color online) The solid lines represent the electron dispersions (\ref{eq6})
  along the $K'-\Gamma-K$ line in the Brillouin zone, for $t = 2.52$ eV. 
  The dashed lines are the asymmetric dispersions corresponding to the finite overlap integral
  $s=0.07$ and  $t = 3$ eV.
  The diamonds are the dispersions obtained by solving the LDA--Kohn--Sham equations. \cite{Despoja13}
  The dot-dashed line labels the position of the Fermi level $E_{\rm F}$ in a typical hole-doped case
  ($E_{\rm F} = -0.5$ eV). \cite{Novoselov05,Zhang05,Li08,Castro09}
  }
  \end{figure}

Here, $t_1=t_2=t_3 \equiv t$ are the bond energies in equilibrium, associated with electron hopping
processes from the $2p_z$ orbital in question to three neighboring carbon atoms
at positions  ${\bf r}_1 = ({\bf a}_1 + {\bf a}_2)/3$, 
${\bf r}_2 = ({\bf a}_2 -2 {\bf a}_1)/3$, and ${\bf r}_3 = ({\bf a}_1 -2 {\bf a}_2)/3$
[${\bf a}_1=a(1,0)$ and ${\bf a}_2 = a(1/2, \sqrt{3}/2)$ are the primitive vectors 
of the Bravais lattice and $a = \sqrt{3} a_{\rm CC} = 2.46$ \AA ].
	The electron dispersions (\ref{eq6}) are illustrated  in Fig.~1 by the solid lines, while the diamonds represent 
the dispersions from Ref.~\onlinecite{Despoja13} obtained by solving the {\it ab initio} LDA--Kohn--Sham equations. 

A more realistic tight-binding model includes the overlap between the neighboring 
$2p_z$ orbitals, described by the overlap parameter $s$, and/or the hopping between second neighbors, described by the parameter $t'$. \cite{Reich02,Castro09}
	In the $t'=0$ case, the resulting electron dispersions are of the form 
$\varepsilon_{\pi^*} ({\bf k}) =  |t ({\bf k})|/(1- (s/t) |t ({\bf k})|)$ and
$\varepsilon_{\pi} ({\bf k}) =  -|t ({\bf k})|/(1 + (s/t) |t ({\bf k})|)$ (dashed lines in the figure).
	The comparison with the LDA-Kohn--Sham dispersions shows that $t \approx 3$ eV and $s \approx 0.07$.
	Without loss of generality, here we restrict our attention to the $s=0$, $t'=0$ case,
with $t \approx 2.52$ eV,  where all relevant vertex functions in $H$ are simple functions of the auxiliary phase $\theta_{\bf k}$ (see Appendix C) and
the effective mass parameter $m_{xx} = (2 \hbar^2/t a^2)$ is equal to the free electron mass.
	As seen in the figure, this tight-binding dispersions give a reasonable approximation for occupied electronic 
states in the hole-doped case.

The coupling between conduction electrons and external electromagnetic fields is obtained by the gauge-invariant 
tight-binding minimal substitution. \cite{Schrieffer64,Peres06,Ziegler06,Stauber08,Lewkowicz09,Kupcic13}
	The result is the expression (\ref{eqB1}) in Appendix B.
	However, for the longitudinal polarization of the fields, the case which is of primary interest
here, we can use the gauge
${\bf E} ({\bf r},t) = -\partial V^{\rm tot} ({\bf r},t)/\partial {\bf r}$ and write the coupling Hamiltonian as
\begin{eqnarray}
&& \hspace{-10mm}
H^{\rm ext}  =  \sum_{{\bf q}}  V^{\rm ext} ({\bf q}) \hat \rho (-{\bf q}),
\label{eq8}
 \end{eqnarray} 
where
\begin{eqnarray}
\hat \rho ({\bf q}) = \sum_{LL'}\sum_{{\bf k}\sigma} e q^{LL'} ({\bf k}, {\bf k}_+)
c^\dagger_{L{\bf k} \sigma} c_{L'{\bf k}+{\bf q} \sigma}
\label{eq9}
\end{eqnarray}
is the total monopole-charge density operator, consisting of the intraband ($L'=L$) and interband 
($L'\neq L$) contributions, and ${\bf k}_+={\bf k}+{\bf q}$.
	The general structure of the monopole-charge vertex functions  $q^{LL'} ({\bf k}, {\bf k}_+)$, 
as well as of the corresponding current vertex functions $J_\alpha^{LL'} ({\bf k}, {\bf k}_+)$, 
is given in Appendix B as well.
	We will also see in Appendix A that there is a close relation between
these two vertex functions, Eq.~(\ref{eqA3}), in which the wave vector ${\bf q}$ can take any direction. \cite{Kupcic13} 
	In the simplest case, corresponding to ${\bf q} = q_\alpha \hat e_\alpha$, this relation reduces to
\begin{eqnarray}
\hbar q_\alpha J_\alpha^{LL'} ({\bf k},{\bf k}_+) = 
\varepsilon_{L'L} ({\bf k}_+,{\bf k}) e q^{LL'} ({\bf k}, {\bf k}_+),
\label{eq10}
\end{eqnarray}
where $\varepsilon_{L'L}({\bf k}',{\bf k})=\varepsilon_{L'}({\bf k}') - \varepsilon_{L}({\bf k})$.

$H^{\rm ph}_0$ is the bare phonon Hamiltonian
\begin{eqnarray}
&& \hspace{-10mm}
H^{\rm ph}_0 = \sum_{\nu {\bf q}'} \frac{1}{2M_\nu} \big[  p^{\dagger}_{\nu {\bf q}'} p_{\nu {\bf q}'} + \big( 
M_\nu \omega_{\nu {\bf q}'} \big)^2 u^{\dagger}_{\nu {\bf q}'} u_{\nu {\bf q}'}\big]
\label{eq11}
 \end{eqnarray} 
given in terms of the phonon field $u_{\nu {\bf q}'}$ and the conjugate field $p_{\nu {\bf q}'}$.
	Here, $\omega_{\nu {\bf q}'}$ is the bare phonon frequency, $\nu$ is the phonon branch index, 
and $M_\nu$ is the corresponding effective ion mass.
	The electron-phonon coupling Hamiltonian can be shown in the following way
\begin{eqnarray}
&& \hspace{-5mm}
H'_1 =  \sum_{\nu LL'} \sum_{{\bf k} {\bf q}' \sigma} \frac{G_\nu^{L'L} ({\bf k}_+,{\bf k})}{
\sqrt{N}} \big(b_{\nu {\bf q}'} + b^\dagger_{\nu -{\bf q}'} \big)
c^{\dagger}_{L'{\bf k}+{\bf q}' \sigma} c_{L{\bf k} \sigma}
\nonumber \\
&& \hspace{0mm}
\equiv  \sum_{\nu {\bf q}'} \frac{g_\nu}{\sqrt{N}} u_{\nu {\bf q}'} \sum_{LL'} \sum_{{\bf k} \sigma} 
q_\nu^{L'L} ({\bf k}_+,{\bf k}) 
c^{\dagger}_{L'{\bf k}+{\bf q}' \sigma} c_{L{\bf k} \sigma},
\label{eq12}
\end{eqnarray} 
where $u_{\nu {\bf q}'} = \sqrt{(\hbar/2M_\nu \omega_{\nu {\bf q}'})} (b_{\nu {\bf q}'} + b^\dagger_{\nu -{\bf q}'} )$
and ${\bf k}_+ = {\bf k} + {\bf q}'$.
	This expression includes the scattering by acoustic and optical phonons as well as the scattering 
by static disorder.
	The latter scattering channel will be labeled by $\nu = 0$.
	For example, to obtain the corresponding $(H_1')^2$ contribution to the memory function 
$M_\alpha({\bf k},\omega)$ in Eq.~(\ref{eq68}), we set the frequency $\omega_{0{\bf q}'}$ equal to zero
and replace $|G_0^{LL'}({\bf k},{\bf k}_+)|^2 [1+2 f^b (\omega_{0 {\bf q}'})]/N$ by 
$|V^{LL'}({\bf k},{\bf k}_+)|^2$
[$V^{LL}({\bf k},{\bf k}_+)=(V({\bf q})/N) \sum_l U_{\bf k}(l,L) U^*_{{\bf k}+{\bf q}} (l,L')$ 
is the usual parameterization of the intraband scattering term \cite{Peres08}].
	The coupling between conduction electrons and in-plane optical phonons in graphene is described by
$q_{\nu }^{LL'} ({\bf k},{\bf k}_+)$, which is given by Eq.~(\ref{eqC12}). \cite{Ando06,Castro07,Kupcic12}

In the short-range part of $H'_2$, it is common to use the intraband scattering approximation, 
\cite{Dzyaloshinskii73,Solyom79}
where the scattering processes in which electrons change the band are neglected, resulting in
\begin{eqnarray}
&& \hspace{-2mm}
H'_2 =  \frac{1}{2V} \sum_{LL'L_1L_1'} \sum_{{\bf k}{\bf k}'{\bf q}} \sum_{\sigma \sigma'} 
\varphi_{\sigma \sigma'}^{LL_1'L_1L'} ({\bf q}) c^{\dagger}_{L{\bf k} \sigma}
c^{\dagger}_{L_1'{\bf k}'+{\bf q}\sigma'}
\nonumber \\
&& \hspace{7mm} \times 
c_{L_1{\bf k}'\sigma'} c_{L'{\bf k}+{\bf q} \sigma}
\nonumber \\
&& \hspace{5mm}
=  \frac{1}{2V} \hspace{-2mm} \sum_{LL'L_1L_1'} \sum_{{\bf q} \sigma \sigma'} 
\varphi_{\sigma \sigma'}^{LL_1'L_1L'} ({\bf q})
\hat \rho^{LL'}_\sigma ({\bf q}) \hat \rho^{L_1'L_1}_{\sigma'} (-{\bf q}).
\nonumber \\
\label{eq13}
\end{eqnarray} 
	The bare Coulomb interaction 
$\varphi_{\sigma \sigma'}^{LL_1'L_1L'} ({\bf q}) \approx e^2 v({\bf q}) 
+ \delta_{L',L} \delta_{L_1,L} \delta_{L_1',L}g_{\sigma \sigma'} ({\bf q})$, is decomposed into
the long-range Coulomb term $v({\bf q})$ ($= 2 \pi/q$) and into
the total intraband short-range interaction $g_{\sigma \sigma'} ({\bf q})$.

\section{Generalized Kubo formulae}
In the microscopic gauge-invariant analysis of the conductivity tensor $\sigma_{\alpha \alpha} ({\bf q},\omega)$ 
in the case in which local field effects can be neglected, it is convenient to use the four-current 
representation of the density operators $\hat J_\mu ({\bf q})$ and introduce 
the microscopic real-time RPA irreducible $4 \times 4$ response tensor by \cite{Schrieffer64,Kubo95}
\begin{eqnarray}
&& \hspace{-11mm}
V \pi_{\mu \nu}({\bf q},t) = -\frac{{\it i}}{\hbar} \theta(t) 
\big< \big[ \hat J_\mu ({\bf q},t), \hat J_\nu (-{\bf q},0) \big] \big>_{\rm irred}
\nonumber \\
&& \hspace{6mm}
\equiv \langle \langle \hat J_\mu ({\bf q}) ; \hat J_\nu (-{\bf q}) \rangle \rangle_t^{\rm irred}
\nonumber \\
&& \hspace{6mm}
\equiv \theta(t) \Psi_{\mu \nu}({\bf q},t).
\label{eq14}
\end{eqnarray}
	The density operators are given by
\begin{eqnarray}
\hat J_\mu ({\bf q}) = \sum_{LL'}\sum_{{\bf k}\sigma} J_\mu^{LL'} ({\bf k}, {\bf k}_+)
c^\dagger_{L{\bf k} \sigma} c_{L'{\bf k}+{\bf q} \sigma},
\label{eq15}
\end{eqnarray}
with 
\begin{eqnarray}
J^{LL'}_\mu ({\bf k}, {\bf k}_+) = 
\left\{ \begin{array}{ll}
J^{LL'}_\alpha ({\bf k}, {\bf k}_+), & \hspace{2mm} \mu = \alpha = 1, 2, 3 \\
& \\
e q^{LL'}({\bf k},{\bf k}_+), &  \hspace{2mm} \mu = 0
\end{array} \right. .
\label{eq16}
\end{eqnarray}
		The $\mu = \alpha =  x, y, z$ are the three current vertices and 
$\mu =0$ is the monopole-charge vertex function from Eq.~(\ref{eq9}).
	The band index $L$ runs over all bands of interest.

It is customary to show the Fourier transform of $\pi_{\mu \mu} ({\bf q}, t)$ as the Fourier--Laplace transform 
of the response function $\Psi_{\mu \mu}({\bf q},t)$, \cite{Kubo95}
\begin{eqnarray}
&& \hspace{-11mm}
V \pi_{\mu \mu}({\bf q},\omega) 
= \int_{0}^\infty {\rm d} t\, {\rm e}^{{\it i} \omega t - \eta t}  \Psi_{\mu \mu}({\bf q},t).
\label{eq17}
\end{eqnarray}
	This expression can be integrated by parts with respect to time twice, leading to
\begin{eqnarray}
&& \hspace{-11mm}
V \pi_{\mu \mu}({\bf q},\omega) = -\frac{1}{\omega^2} \big[ \Phi_{\mu \mu}(\omega) - \Phi_{\mu \mu}(0) \big],
\label{eq18}
\end{eqnarray}
where
\begin{eqnarray}
&& \hspace{-11mm}
\Phi_{\mu \mu}(\omega) = \langle \langle [\hat J_\mu ({\bf q}),H] ; [\hat J_\mu (-{\bf q}),H] \rangle \rangle_\omega^{\rm irred}.
\label{eq19}
\end{eqnarray}
	The expressions (\ref{eq18})--(\ref{eq19}) will be referred to as the generalized Kubo formulae 
for the four-current correlation functions $\pi_{\mu \mu}({\bf q},\omega)$.
	Their importance is twofold.

For $\mu=0$, it is easily seen that the commutator in Eq.~(\ref{eq19}) is actually the definition relation 
for the current density operator $\hat J_\alpha ({\bf q})$,
\begin{eqnarray}
&& \hspace{-10mm}
[\hat J_0 ({\bf q}),H] \approx [\hat J_0 ({\bf q}),H_0^{\rm el}] = \sum_\alpha \hbar q_\alpha \hat J_\alpha ({\bf q}).
\label{eq20}
\end{eqnarray}
	In this case, Eqs.~(\ref{eq18}) and (\ref{eq19}) reduce to the well-known results, the first and the second 
Kubo formula for the conductivity tensor \cite{Kubo95}
\begin{eqnarray}
&& \hspace{-10mm}
\sum_{\beta}  \sigma_{\alpha \beta} ({\bf q}, \omega) q_\beta = {\it i} \pi_{\alpha 0} ({\bf q}, \omega),
\label{eq21} \\
&& \hspace{-1mm}
\sigma_{\alpha \beta} ({\bf q}, \omega) =  \frac{{\it i}}{\omega}  
\big[ \pi_{\alpha \beta} ({\bf q}, \omega)-\pi_{\alpha \beta} ({\bf q})\big]. 
\label{eq22}
\end{eqnarray}
	For $\mu=\alpha$, on the other hand, Eqs.~(\ref{eq18}) and (\ref{eq19}) give the basic relations from 
the microscopic memory-function theory. \cite{Gotze72}
	These expressions will be studied in detail in Ref.~\onlinecite{KupcicUP}.
	In the present two-band case,  Eqs.~(\ref{eq21}) and (\ref{eq22}) reduce to
\begin{eqnarray}
&& \hspace{-10mm}
\sum_{\beta}  \sigma_{\alpha \beta}^{\rm tot} ({\bf q}, \omega) q_\beta = {\it i} \pi_{\alpha 0}^{\rm tot} ({\bf q}, \omega),
\label{eq23} \\
&& \hspace{-1mm}
\sigma_{\alpha \beta}^{\rm tot} ({\bf q}, \omega) =  \frac{{\it i}}{\omega}  
\big[ \pi_{\alpha \beta}^{\rm tot} ({\bf q}, \omega)-\pi_{\alpha \beta}^{\rm tot} ({\bf q})\big]. 
\label{eq24}
\end{eqnarray}

\begin{figure}
   \centerline{\includegraphics[width=20pc]{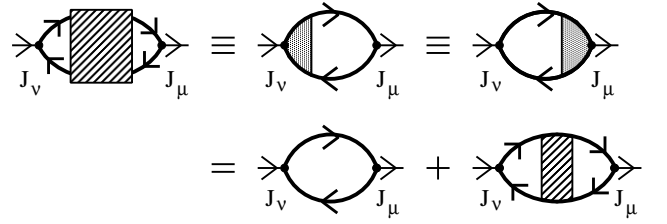}}
   \caption{The Bethe--Salpeter expression for the current-current correlation function
   $\pi_{\mu \nu}^{\rm C} (q)$. \cite{Schrieffer64,Dzyaloshinskii73,Kupcic13}
   }
  \end{figure}

\section{Ward identity}
To understand the way in which the vertex corrections enter in the conductivity tensor within the relaxation-time approximation,
it is helpful also to derive the relations (\ref{eq21}) and (\ref{eq22}) at zero temperature beginning with
the definition of the causal RPA irreducible $4 \times 4$ response tensor
\cite{Abrikosov75,Fetter71,Mahan90}
\begin{eqnarray}
&& \hspace{-11mm}
\hbar V \pi^{\rm C}_{\mu \nu}({\bf q},t) = -{\it i} 
\frac{\big< \Psi_0 \big| T \big[ \hat J_\mu ({\bf q},t) \hat J_\nu (-{\bf q},0) \big] \big| \Psi_0 \big>_{\rm irred}}{\big< \Psi_0 \big| \Psi_0 \big>} .
\nonumber \\
\label{eq25}
\end{eqnarray}
	To do this, we first use the usual definition of the auxiliary RPA irreducible electron-hole propagator
\cite{Schrieffer64,Kupcic13}
\begin{eqnarray}
\hbar^2 \Lambda^{LL'}_\mu (k,k_+) = {\cal G}_L (k) {\cal G}_{L'} (k_+) \Gamma^{LL'}_{\mu}(k,k_+)
\label{eq26}
\end{eqnarray}
and remember that $\pi^{\rm C}_{\mu \nu}({\bf q},\omega)$ can be expressed in terms of $\Lambda^{LL'}_\mu (k,k_+)$ 
in the following two equivalent ways
\begin{eqnarray}
&& \hspace{-5mm}
\pi^{\rm C}_{\mu \nu}(q) = -{\it i} \hbar \sum_{LL'} \sum_\sigma \int \frac{{\rm d}^{d+1} k}{(2 \pi)^{d+1}} \, 
J^{LL'}_{\mu}({\bf k},{\bf k}_+) \Lambda^{L'L}_\nu (k_+,k) 
\nonumber \\
\label{eq27}
\end{eqnarray}
and
\begin{eqnarray}
&& \hspace{-5mm}
\pi^{\rm C}_{\mu \nu}(q) = -{\it i} \hbar \sum_{LL'} \sum_\sigma \int \frac{{\rm d}^{d+1} k}{(2 \pi)^{d+1}} \,  
\Lambda^{LL'}_\mu (k,k_+)J^{L'L}_\nu ({\bf k}_+,{\bf k})
\nonumber \\
\label{eq28}
\end{eqnarray}
(see Fig.~2).
	Equations (\ref{eq27}) and (\ref{eq28}) are known as the Bethe--Salpeter expressions for 
$\pi^{\rm C}_{\mu \nu}(q)$. 
	In Eq.~(\ref{eq26}), $\Gamma^{LL'}_{\mu}(k,k_+)$ is a renormalized version of the vertex function 
$J_\mu^{LL'} ({\bf k},{\bf k}_+)$, and $k = ({\bf k}, k_0)$, with $k_0= \omega$, is the four-component wave vector.
	Finally, $d=2$ in graphene.

The Ward identity is the identity relation connecting $\Gamma_0^{LL'} (k,k_+) $ 
with the three renormalized current vertices $\Gamma_\alpha^{LL'} (k,k_+)$. 
\cite{Schrieffer64}
	The straightforward calculation leads to
\begin{eqnarray}
&& \hspace{-10mm}
\sum_{\mu= 0}^3 q_\mu \Gamma^{LL'}_\mu (k,k_+) =
\sum_{\alpha} q_{\alpha} \Gamma^{LL'}_\alpha (k,k_+) - \omega \Gamma^{LL'}_0 (k,k_+)
\nonumber \\
&& \hspace{18mm}
= J_0^{LL'}({\bf k},{\bf k}_+) \big[{\cal G}^{-1}_L (k) - {\cal G}^{-1}_{L'} (k_+)\big].
\nonumber \\
\label{eq29}
\end{eqnarray}
	It can also be shown in the following way
\begin{eqnarray}
&& \hspace{-5mm}
\sum_{\mu= 0}^3 q_\mu \hbar ^2 \Lambda^{LL'}_\mu (k,k_+) = J_0^{LL'}({\bf k},{\bf k}_+) 
\big[{\cal G}_{L'}(k_+)-{\cal G}_L (k)\big].
\nonumber \\
\label{eq30}
\end{eqnarray}
	The difference ${\cal G}^{-1}_L (k) - {\cal G}^{-1}_{L'} (k_+)$ on the right-hand side 
of Eq.~(\ref{eq30}) satisfies the Dyson relation
\begin{eqnarray}
&& \hspace{-10mm}
{\cal G}^{-1}_L (k) - {\cal G}^{-1}_{L'} (k_+)
\nonumber \\
&& \hspace{0mm}
\approx \varepsilon_{L'L} ({\bf k}_+,{\bf k})/\hbar - \omega + \Sigma_{L'}(k_+) - \Sigma_L(k),
\label{eq31}
\end{eqnarray}
with $\varepsilon_{LL}({\bf k}_+,{\bf k})/\hbar -\omega \approx 
\sum_{\mu} q_\mu v^{L,0}_\mu ({\bf k},{\bf k}_+)$ in the intraband channel.
	The relation (\ref{eq29}) is the generalization of the well-known single-band Ward identity 
\cite{Schrieffer64}
to the multiband case.
	Not surprisingly, in the ideal conductivity regime it reduces to Eqs.~(\ref{eq10}) and (\ref{eqA4}).
	Notice that in this case the factor $q_\alpha$ on the right-hand side of equation comes from the expansion of 
$\varepsilon_{LL}({\bf k}_+,{\bf k})$ and $q^{LL'} ({\bf k}, {\bf k}_+)$, $L' \neq L$, in powers of $q_\alpha$.

After simple algebraic manipulations with Eqs.~(\ref{eq28}) and (\ref{eq30}), we obtain the relation
\begin{eqnarray}
\sum_\mu q_\mu \pi^{\rm C}_{\mu \nu}(q) =
\sum_\mu q_\mu \pi_{\mu \nu}(q) = - \sum_\mu q_\mu \frac{e^2 n_{\mu \nu}({\bf q})}{m}.
\label{eq32}
\end{eqnarray}
	The latter is known as the four-current representation of the charge continuity equation,
which takes care of both local charge conservation and gauge invariance.
	In this expression, $n_{0\nu}({\bf q})=n_{\nu 0}({\bf q})= 0$,
\begin{eqnarray}
&& \hspace{-5mm} n_{\alpha \beta}({\bf q}) = \sum_{LL'}
\frac{1}{V} \sum_{{\bf k} \sigma} \frac{m}{e^2}
\frac{J_\alpha^{LL'} ({\bf k},{\bf k}_+) 
J_{\beta}^{L'L} ({\bf k}_+,{\bf k})}{\varepsilon_{L'L} ({\bf k}_+,{\bf k})} 
\nonumber \\
&& \hspace{10mm} \times 
\big[n_L({\bf k}) - n_{L'}({\bf k}_+) \big]
\label{eq33}
\end{eqnarray}
is the total number of charge carriers, and
\begin{eqnarray}
n_L({\bf k}) = -{\it i} \int_{-\infty}^\infty \frac{{\rm d} k_0}{2 \pi} {\cal G}_L (k)
\label{eq34}
\end{eqnarray}
is the momentum distribution function at zero temperature.
	This quantity is found to be essential for understanding the electrodynamic properties of quasi-one-dimensional
systems \cite{Dzyaloshinskii73}
as well as the ballistic conductivity regime in graphene \cite{Peres06}.

The effective number of charge carriers $n^{\rm intra}_{\alpha \beta}$, defined by 
\begin{eqnarray}
&& \hspace{-5mm} n^{\rm intra}_{\alpha \beta} = \frac{1}{V} \sum_{L{\bf k} \sigma} m 
v_\alpha^{L} ({\bf k}) v_\beta^{L} ({\bf k})
\bigg( -\frac{\partial n_L({\bf k})}{\partial \varepsilon_L({\bf k})}\bigg)
\nonumber \\
&& \hspace{4mm} 
= \frac{1}{V} \sum_{L{\bf k} \sigma} \gamma_{\alpha \beta}^{LL} ({\bf k}) n_L({\bf k}),
\label{eq35}
\end{eqnarray}
and
\begin{eqnarray}
&& \hspace{-5mm} n^{\rm inter}_{\alpha \beta} = \sum_{L' (\neq L)}
\frac{1}{V} \sum_{L{\bf k} \sigma}\frac{m}{e^2} \bigg(
\frac{J_\alpha^{LL'} ({\bf k},{\bf k}_+) J_{\beta}^{L'L} ({\bf k}_+,{\bf k})}{
\varepsilon_{L'L} ({\bf k}_+,{\bf k})} 
\nonumber \\
&& \hspace{7mm} 
+ \frac{J_\alpha^{L'L} ({\bf k}_+,{\bf k}) J_{\beta}^{LL'} ({\bf k},{\bf k}_+)}{
\varepsilon_{L'L} ({\bf k}_+,{\bf k})} \bigg) n_L({\bf k})
\label{eq36}
\end{eqnarray}
are, respectively, the intraband and interband parts in $n_{\alpha \beta}({\bf q})$ at ${\bf q} \approx 0$.
	Here, $\gamma_{\alpha \beta}^{LL} ({\bf k}) = 
(m/\hbar^2) \partial^2 \varepsilon_L ({\bf k}) /\partial k_\alpha \partial k_\beta$ 
is the dimensionless reciprocal effective mass tensor [in graphene, it is given by Eq.~(\ref{eqC9})].

The expression (\ref{eq32}) represents a compact way of writing the relations \cite{Schrieffer64}
\begin{eqnarray}
&& \hspace{-5mm}
\omega \pi_{0 0}(q) = \sum_\alpha q_\alpha \pi_{\alpha 0}(q), 
\label{eq37} \\
&& \hspace{-5mm}
\omega \pi_{0 \alpha}(q) = \sum_{\beta}q_{\beta} \bigg(\pi_{\beta \alpha}(q)  
+ \frac{e^2 n_{\beta \alpha}({\bf q})}{m}\bigg).
\label{eq38}
\end{eqnarray}
	In the normal metallic state, Eq.~(\ref{eq38}) is nothing more than the relation (\ref{eq22}), because 
\begin{eqnarray}
\frac{e^2}{m} n_{\beta \alpha}({\bf q}) = -\pi_{\beta \alpha}({\bf q})
\label{eq39}
\end{eqnarray}
in this case.
	Similarly, Eq.~(\ref{eq37}), together with Eq.~(\ref{eq21}), gives the gauge-invariant 
form of the dielectric susceptibility \cite{Kubo95}
\begin{eqnarray}
&& \hspace{-10mm}
\chi ({\bf q}, \omega) \equiv \pi_{00} ({\bf q}, \omega) 
= \frac{1}{{\it i}\omega} \sum_{\alpha \beta} q_\alpha \sigma_{\alpha \beta} ({\bf q}, \omega) q_\beta,
\label{eq40}
\end{eqnarray}
which is consistent with the aforementioned definition of the macroscopic dielectric function, Eq.~(\ref{eq1}).
	In the present case, this expression reduces to 
\begin{eqnarray}
&& \hspace{-10mm}
\chi^{\rm tot} ({\bf q}, \omega) \equiv \pi_{00}^{\rm tot} ({\bf q}, \omega) 
= \frac{1}{{\it i}\omega} \sum_{\alpha \beta} q_\alpha \sigma_{\alpha \beta}^{\rm tot} ({\bf q}, \omega) q_\beta.
\label{eq41}
\end{eqnarray}

\section{Intraband dynamical conductivity}
\subsection{Hydrodynamic formulation}
An essential step towards the general microscopic formulation of electrodynamic properties of 
multiband electronic systems is to separate 
the intraband contributions to the microscopic response functions from the interband ones.
	In most cases of interest the low-energy physics is completely described in terms 
of the intraband contributions, and in a rich variety of weakly interacting electronic systems 
we can introduce the quantity usually called the intraband memory function $M^{LL} ({\bf k},{\bf q},\omega)$ phenomenologically 
and describe the macroscopic response functions in question in terms of $M^{LL} ({\bf k},{\bf q},\omega)$. \cite{Kubo95,Forster75}
	In the diagrammatic language, the memory function $M^{LL} ({\bf k},{\bf q},\omega)$ 
is nothing but the self-energy of the intraband electron-hole pair in the approximation called the memory-function 
approximation. \cite{Kupcic13,KupcicUP}
	In the case in which $M^{LL} ({\bf k},{\bf q},\omega)$ is independent of $\omega$, the memory function reduces 
to the relaxation rate $\Gamma^{LL}_\alpha ({\bf k})$ multiplied by ${\it i}$; i.e., 
$M^{LL} ({\bf k},{\bf q}) \approx {\it i} \Gamma^{LL}_\alpha ({\bf k}) \equiv {\it i} /\tau^{L}_{\rm tr} ({\bf k})$.
	Therefore, to obtain the intraband memory-function conductivity formula in a phenomenological way,
it usually suffices to use the common textbook form \cite{Pines89,Ziman72,Abrikosov88}
of the intraband conductivity obtained by means of the relaxation-time approximation 
and replace ${\it i}/\tau_{\rm tr} ({\bf k})$ by $M^{LL} ({\bf k},{\bf q},\omega)$.

	Caution is in order regarding the ballistic conductivity regime in graphene where the interband conductivity 
is non-zero down to $\omega \approx 0$.
	For this reason, the general expressions presented below are expected to be directly applicable to doped graphene 
for $|E_{\rm F}|$ not too small.
	In the $|E_{\rm F}| \rightarrow 0$ limit, the result depends on how the $\omega \rightarrow 0$ limit is taken, 
as already pointed out in Refs.~\onlinecite{Ziegler06} and \onlinecite{Lewkowicz09}.

To obtain a rough justification of this simple method of calculating $\sigma_{\alpha \alpha}^{\rm intra} ({\bf q}, \omega)$ beyond the relaxation-time approximation, let us consider the common hydrodynamic derivation of this function.
	Our puprose here is to present the formalism which includes the intraband electron-electromagnetic field
vertex corrections in a natural way, at variance with the response theory \cite{Castro09,Peres08,Carbotte10}
usually used in graphene in which these corrections are neglected.
	Evidently it is not easy to accept the quantitative description of the low-energy physics in both pure 
and doped graphene within the response theory in which the leading role is played by the ${\bf q} \approx 0$ scattering processes and, at the same time, the electron-electromagnetic field vertex corrections, which lead to the identical cancellation of these scattering processes, are disregarded.

We combine here the constitutive relation for the microscopic real-time RPA irreducible current-monopole correlation function $\pi_{\alpha 0}^{\rm intra} ({\bf q}, \omega)$ from Eq.~(\ref{eq21}), 
\begin{eqnarray}
&& \hspace{-10mm}
J_\alpha^{\rm intra} ({\bf q}, \omega) = \pi_{\alpha 0}^{\rm intra} ({\bf q}, \omega) V^{\rm tot} ({\bf q},\omega)
\nonumber \\
&& \hspace{8mm}
= \frac{1}{V} \sum_{L{\bf k}\sigma} 
J_\alpha^{LL}({\bf k},{\bf k}_+) \langle c^\dagger_{L{\bf k} \sigma} c_{L{\bf k}_+{\bf q} \sigma} \rangle_\omega
\nonumber \\
&& \hspace{8mm}
\approx \frac{1}{V} \sum_{L{\bf k}\sigma} 
e v_\alpha^{L}({\bf k}) \delta n^{LL} ({\bf k}, {\bf q}, \omega),
\label{eq42}
\end{eqnarray}
with the generalized self-consistent RPA equation
\begin{eqnarray}
&& \hspace{-5mm}
\bigg({\it i} \hbar \frac{\partial}{\partial t} + \varepsilon_L({\bf k}) - \varepsilon_L({\bf k}_+) \bigg)
\big(c^\dagger_{L{\bf k} \sigma} c_{L{\bf k}+{\bf q} \sigma} \big)_t
\nonumber \\
&& \hspace{2mm}
= - \int_{-\infty}^t {\rm d} t' \, \hbar M^{LL} ({\bf k},{\bf q},t-t') 
\big(c^\dagger_{L{\bf k} \sigma} c_{L{\bf k}+{\bf q} \sigma} \big)_{t'}
\nonumber \\
&& \hspace{5mm}
+ \big( [c^\dagger_{L{\bf k} \sigma} c_{L{\bf k}+{\bf q} \sigma},H] \big)_t^{\rm stoh}
+ \big( c^\dagger_{L{\bf k} \sigma} c_{L{\bf k}\sigma}
\nonumber \\
&& \hspace{5mm}
- c^\dagger_{L{\bf k}+{\bf q} \sigma} c_{L{\bf k}+{\bf q} \sigma} \big)
e q^{LL}({\bf k}_+,{\bf k})V^{\rm tot} ({\bf q},t).
\label{eq43}
\end{eqnarray}
	Here, 
$V^{\rm tot} ({\bf q},t)=V^{\rm ext} ({\bf q},t)+V^{\rm ind} ({\bf q},t) =({\it i}/q_\alpha) E_\alpha({\bf q},t)$ 
is the RPA screened scalar potential, and $E_\alpha({\bf q},t)$ is the corresponding macroscopic electric field.
	The expression in the third row of Eq.~(\ref{eq42}) is the standard Fermi liquid representation for 
$J_\alpha^{\rm intra} ({\bf q}, \omega)$, \cite{Pines89}
where 
$n^{LL} ({\bf k}, {\bf q}, \omega) = n_L({\bf k}) + \delta n^{LL} ({\bf k}, {\bf q}, \omega) = n_L({\bf k}) +
\langle c^\dagger_{L{\bf k} \sigma} c_{L{\bf k}_+{\bf q} \sigma} \rangle_\omega$ and
$v_\alpha^L({\bf k})=(1/\hbar) \partial \varepsilon_L ({\bf k}) / \partial k_\alpha$ 
represent, respectively, the non-equilibrium distribution function and the bare electron group velocity.
	Finally, $n_L({\bf k})=\langle c^\dagger_{L{\bf k} \sigma} c_{L{\bf k} \sigma} \rangle$ 
is the momentum distribution function.

The equation (\ref{eq43}) is reminiscent of the generalized Langevin equation in which 
$\big( [c^\dagger_{L{\bf k} \sigma} c_{L{\bf k}+{\bf q} \sigma},H] \big)_t^{\rm stoh}$ 
plays the role of the stochastic force and the term containing $M^{LL} ({\bf k},{\bf q},t-t')$  is the friction term.
	It is easy to draw standard conclusions from this equation. \cite{Kubo95,Forster75}
	After performing a Fourier transformation in time, the equation for the non-equilibrium average of
$\big(c^\dagger_{L{\bf k} \sigma} c_{L{\bf k}+{\bf q} \sigma} \big)_\omega$ becomes
\begin{eqnarray}
&& \hspace{-5mm}
\big[ \hbar \omega +  \hbar M^{LL} ({\bf k},{\bf q},\omega) + \varepsilon_L({\bf k}) - \varepsilon_L({\bf k}_+) \big]
\delta n^{LL} ({\bf k}, {\bf q}, \omega)
\nonumber \\
&& \hspace{5mm}
= \big[ n_L({\bf k}) - n_L({\bf k}_+) \big] e q^{LL}({\bf k}_+,{\bf k}) V^{\rm tot} ({\bf q},\omega).
\nonumber \\
\label{eq44}
\end{eqnarray}
	The result is the expression for the intraband conductivity tensor
$\sigma_{\alpha \alpha}^{\rm intra} ({\bf q}, \omega) = ({\it i}/q_\alpha) \pi_{\alpha 0}^{\rm intra} ({\bf q}, \omega)$
[the intraband part in Eq.~(\ref{eq23})],
\begin{eqnarray}
&& \hspace{-5mm} 
\sigma_{\alpha \alpha}^{\rm intra} ({\bf q}, \omega) =
\frac{1}{V} \sum_{L{\bf k} \sigma}
{\it i} \hbar |J^{LL}_{\alpha}({\bf k},{\bf k}_+)|^2 
\frac{n_L({\bf k})-n_L({\bf k}_+)}{\varepsilon_{LL}({\bf k}_+,{\bf k})}
\nonumber \\
&& \hspace{10mm} \times 
\frac{1}{\hbar \omega + \varepsilon_{LL}({\bf k},{\bf k}_+)
+  \hbar M^{LL} ({\bf k},{\bf q},\omega)},
\label{eq45}
\end{eqnarray}
which covers all physically relevant regimes with the exception of the static screening.

On the other hand, the standard Fermi liquid theory treats $\sigma_{\alpha \alpha}^{\rm intra} ({\bf q}, \omega)$ 
in a way  consistent with Eq.~(\ref{eq37}).
	It is easily seen that it gives the correct description of the static screening as well. \cite{Pines89,Platzman73}
	In this case, Eq.~(\ref{eq44}) is replaced by 
\begin{eqnarray}
&& \hspace{-10mm}
\big[ \hbar \omega + \varepsilon_L({\bf k}) - \varepsilon_L({\bf k}_+)  \big] \delta n^{LL} ({\bf k}, {\bf q}, \omega)
\nonumber \\
&& \hspace{0mm}
+ \hbar M^{LL} ({\bf k},{\bf q},\omega)\delta n^{LL}_1 ({\bf k}, {\bf q}, \omega)
\nonumber \\
&& \hspace{0mm}
= \big[ n_L({\bf k}) - n_L({\bf k}_+) \big] e q^{LL}({\bf k}_+,{\bf k}) V^{\rm tot} ({\bf q},\omega),
\label{eq46}
\end{eqnarray}
and the result is the following \cite{Kupcic13}
\begin{eqnarray}
&& \hspace{-5mm} 
\sigma_{\alpha \alpha}^{\rm intra} ({\bf q}, \omega) =
\frac{1}{V} \sum_{L{\bf k} \sigma}
{\it i} \hbar |J^{LL}_{\alpha}({\bf k},{\bf k}_+)|^2
\frac{n_L({\bf k})-n_L({\bf k}_+)}{\varepsilon_{LL}({\bf k}_+,{\bf k})}
\nonumber \\
&& \hspace{5mm} \times 
\frac{\hbar \omega}{\hbar \omega(\hbar \omega + 
\hbar M^{LL} ({\bf k},{\bf q},\omega)) - \varepsilon^2_{LL}({\bf k},{\bf k}_+)}.
\label{eq47}
\end{eqnarray}
	As usual, $\delta n^{LL}_1 ({\bf k}, {\bf q}, \omega)$ represents the contribution to 
$\delta n^{LL}({\bf k}, {\bf q}, \omega)$ which is proportional to $v_\alpha^L({\bf k})$, and
$n_L({\bf k}) = (1/\beta \hbar) \sum_{{\it i}\omega_n} {\cal G}_L ({\bf k}, {\it i} \omega_n)$
[this expression for $n_L({\bf k})$ holds in pure graphene as well].
	The same result is obtained in Ref.~\onlinecite{Kupcic13}
by considering the quantum transport equations in the memory-function approximation.

\subsection{Generalized Drude formula}
For long wavelengths, Eqs.~(\ref{eq45}) and (\ref{eq47}) reduce to the macroscopic conductivity tensor from the macroscopic Maxwell equations.
	In this limit, we can use the usual simplifications, 
$J^{LL}_{\alpha}({\bf k},{\bf k}_+) \approx e v_\alpha^L({\bf k})$,
$q^{LL}({\bf k}_+,{\bf k}) \approx 1$,  $\varepsilon_{LL}({\bf k},{\bf k}_+) \approx 0$, and
$M^{LL} ({\bf k},{\bf q},\omega) \approx M_{\alpha}^{LL}({\bf k}, \omega)$
(i.e., the memory function is assumed to depend on the direction of ${\bf q}= \hat e_\alpha q_\alpha$, but not on its magnitude).
	The result is the intraband memory-function conductivity formula
\begin{eqnarray}
&& \hspace{-5mm}
\sigma_{\alpha \alpha}^{\rm intra} (\omega) = \frac{{\it i}e^2}{m} \frac{1}{V} \sum_{L{\bf k} \sigma} 
\bigg(- \frac{\partial n_L({\bf k})}{\partial \varepsilon_L ({\bf k})}\bigg)
\frac{m [v_\alpha^L({\bf k})]^2}{\omega +  M_{\alpha}^{LL}({\bf k}, \omega)}
\nonumber \\
\label{eq48}
\end{eqnarray}
and the expression for the corresponding current-current correlation function 
\begin{eqnarray}
&& \hspace{-5mm}
\pi_{\alpha \alpha}^{\rm intra} (\omega) = \frac{e^2}{m} \frac{1}{V} \sum_{L{\bf k} \sigma} 
m [v_\alpha^L({\bf k})]^2\frac{\partial n_L({\bf k})}{\partial \varepsilon_L ({\bf k})}
\frac{M^{LL}_{\alpha}({\bf k}, \omega)}{\omega +  M^{LL}_{\alpha}({\bf k}, \omega)}.
\nonumber \\
\label{eq49}
\end{eqnarray}
	It is easily seen that the latter function plays an important role in studying the ${\bf q}' \approx 0$ 
in-plane optical phonons in graphene as well. \cite{Ando06}
	It is usually mistaken for the function
\begin{eqnarray}
&& \hspace{-5mm}
\pi_{\alpha \alpha}^{\rm intra} (\omega) - \pi_{\alpha \alpha}^{\rm intra} (0)
\nonumber \\
&& \hspace{5mm}
= - \frac{e^2}{m} \frac{1}{V} \sum_{L{\bf k} \sigma} 
m [v_\alpha^L({\bf k})]^2\frac{\partial n_L({\bf k})}{\partial \varepsilon_L ({\bf k})}
\frac{\omega}{\omega +  M^{LL}_{\alpha}({\bf k}, \omega)}.
\nonumber \\
\label{eq50}
\end{eqnarray}

The generalized Drude formula for conductivity tensor, which is a widely applicable method for analyzing
measured reflectivity spectra, \cite{Uchida91} describes the case in which the dependence of $M_\alpha^{LL} ({\bf k},\omega)$ 
in Eq.~(\ref{eq45}) on ${\bf k}$ and $L$ can be neglected.
	The result is \cite{Gotze72}
\begin{eqnarray}
&& \hspace{-5mm} 
\sigma_{\alpha \alpha}^{\rm intra} (\omega) \approx
\frac{{\it i} e^2 n^{\rm intra}_{\alpha \alpha}}{m(\omega + M_{1\alpha}(\omega))},
\nonumber \\
&& \hspace{-5mm} 
\pi_{\alpha \alpha}^{\rm intra} (\omega) \approx
-\frac{e^2 n^{\rm intra}_{\alpha \alpha}}{m}\frac{M_{1\alpha}(\omega)}{\omega + M_{1\alpha}(\omega)},
\label{eq51}
\end{eqnarray}
where $n^{\rm intra}_{\alpha \alpha}$ is the effective number of charge carriers given by Eq.~(\ref{eq35}).
	For $\omega \ll \Gamma_{1\alpha}(0)$, we can also write
\begin{eqnarray}
&& \hspace{-5mm} 
\sigma_{\alpha \alpha}^{\rm intra} (\omega) \approx
\frac{{\it i} e^2 n^{\rm eff}_{\alpha \alpha}(\omega)}{m(\omega + {\it i} \Gamma_{1\alpha}(\omega))},
\label{eq52}
\end{eqnarray}
where $n^{\rm eff}_{\alpha \alpha}(\omega)
= n^{\rm intra}_{\alpha \alpha}/(1+\lambda_\alpha(\omega))$ is the renormalized effective number of charge carriers, 
$\Gamma_{1 \alpha} (\omega) = M^i_{1 \alpha} (\omega)/(1+\lambda_\alpha(\omega))$, and
$\lambda_\alpha(\omega) = M^r_{1\alpha} (\omega)/\omega$.

\subsection{Ordinary Drude formula}
The ordinary Drude formula follows from Eq.~(\ref{eq52}) after using the relaxation-time approximation, where 
$n^{\rm eff}_{\alpha \alpha}(\omega) \approx n^{\rm eff}_{\alpha \alpha}(0) \equiv n^{\rm eff}_{\alpha \alpha}$ 
and $\Gamma_{1\alpha}(\omega) \approx \Gamma_{1\alpha}(0) \approx \Gamma_{1}$. 
	The result is
\begin{eqnarray}
&& \hspace{-5mm} 
\sigma_{\alpha \alpha}^{\rm intra} (\omega) \approx
\frac{{\it i} e^2 n^{\rm eff}_{\alpha \alpha}}{m(\omega + {\it i} \Gamma_1)}.
\label{eq53}
\end{eqnarray}
	In weakly interacting electronic systems, $\hbar M^i_{1\alpha}(0) = (1+\lambda_\alpha(0))\hbar \Gamma_{1\alpha}$ 
can be extracted from measured dc resistivity data by using the first equality in
\begin{eqnarray}
&& \hspace{-10mm}
\sigma^{\rm dc }_{\alpha \alpha} 
= \frac{e^2 n^{\rm intra}_{\alpha \alpha}}{mM^i_{1 \alpha}(0)}
= \frac{e^2 n^{\rm eff}_{\alpha \alpha}}{m\Gamma_{1 \alpha}} = 
\frac{e^2 n^{\rm intra,0}_{\alpha \alpha}}{mM^{i,0}_{1 \alpha}(0)} = \frac{e^2 n_h}{m \gamma_1}
\label{eq54} 
\end{eqnarray}
[here $V_0n_h = 2-V_0n$, $V_0 = \sqrt{3}a^2/2$ is the unit cell volume, and $n^{\rm intra,0}_{\alpha \alpha}$ is obtained by replacing $n_L({\bf k})$ in Eq.~(\ref{eq35}) by $f_L({\bf k})$].
	In this way, it is possible to get
the damping energy $\hbar M^i_{1\alpha}(0)$ at different temperatures and different doping levels
which gives the exact description of the relaxation processes at zero frequency.
	It is the first important parameter which describes the damping effects in weakly interacting electronic systems.
	Evidently $M^i_{1\alpha}(0)$ must not be confused with $\Gamma_{1\alpha}$.
	It must be noticed that the dc conductivity of hole-doped graphene is usually analyzed by using the expression
\cite{Novoselov05,Zhang05}
\begin{eqnarray}
&& \hspace{-10mm}
\sigma^{\rm dc }_{\alpha \alpha} = e n_h \mu_h,
\label{eq55} 
\end{eqnarray}
where the doped holes are characterized by their mobility $\mu_h = (e/m \gamma_1)$ rather than by the damping energies from Eq.~(\ref{eq54}).

The ${\bf q} \approx 0$ dielectric susceptibility associated with the conductivity (\ref{eq53}) is
\begin{eqnarray}
&& \hspace{-5mm} 
\chi^{\rm intra} ({\bf q},\omega) \approx \sum_{\alpha} 
\frac{q_\alpha^2 e^2 n^{\rm eff}_{\alpha \alpha}}{m\omega (\omega + {\it i} \Gamma_1)}.
\label{eq56}
\end{eqnarray}
	This expression differs from its usual textbook form
\cite{Ziman79,Platzman73,Kupcic00,Hwang07,Despoja13}
\begin{eqnarray}
&& \hspace{-10mm} 
\chi^{\rm intra} ({\bf q}, \omega) =
\frac{e^2}{V} \sum_{L{\bf k} \sigma} \frac{f_L({\bf k})-f_L({\bf k}_+)}{
\hbar \omega + \varepsilon_{LL}({\bf k},{\bf k}_+) +  {\it i} \hbar \Gamma_1}
\nonumber \\
&& \hspace{8mm} 
\approx \sum_{\alpha} 
\frac{q_\alpha^2 e^2 n^{\rm intra,0}_{\alpha \alpha}}{m(\omega + {\it i} \Gamma_1)^2}
\label{eq57}
\end{eqnarray}
by a factor of $(\omega + {\it i} \Gamma_1)/\omega$.

 \begin{figure}[tb]
  \includegraphics[width=15pc]{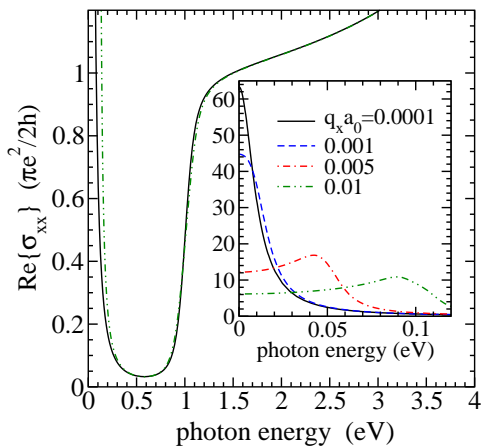}
  \caption{(Color online) The real part of the dynamical conductivity (\ref{eq61}) in hole-doped graphene obtained 
  by means of the relaxation-time approximation.
  The parameters of the model are $t= 2.52$ eV, $E_{\rm F} = -0.5$ eV, $\hbar \Gamma_1 = 10$ meV, 
  $\hbar \Gamma_2 = 50$ meV,  $T=150$ K, and ${\bf q} = (q_x,0)$.
  The solid line in the inset of figure represents the ordinary Drude formula (\ref{eq53}).
  $a_0$ is the Bohr radius.
  }
  \end{figure}

\subsection{Hole-doped graphene}
Actually, there is a wide class of electronic systems (doped graphene being the example) in which the expressions (\ref{eq51})--(\ref{eq54}) and (\ref{eq56}) are applicable.
	Namely, in the case in which the Fermi surface is nearly isotropic and
$M^{LL} ({\bf k},{\bf q},\omega) \approx M_{1\alpha}(|{\bf k}|, \omega)$, we can approximate the memory function
$M^{LL} ({\bf k},{\bf q},\omega)$ by $M_{1\alpha}(k_{\rm F}, \omega)$, and the dynamical conductivity reduces again to Eq.~(\ref{eq51}) with $M_{1\alpha}(\omega)$ replaced by $M_{1\alpha}(k_{\rm F}, \omega)$.
	The dc conductivity and the ${\bf q} \approx 0$ dielectric susceptibility are given, respectively, by
Eqs.~(\ref{eq54}) and (\ref{eq56}) with the implicit dependence of both $M_{1\alpha}(0)$ and $\Gamma_1$ on $k_{\rm F}$.
	The doping-dependent measurements on hole- and electron-doped graphene have shown that 
$M^{i,0}_{1 \alpha}(0) \propto 1/k_{\rm F}$, which, together with $n^{\rm intra,0}_{\alpha \alpha} \propto k_{\rm F}$, leads to the proportionality between $\sigma^{\rm dc }_{\alpha \alpha}$ and $n_h \propto k_{\rm F}^2$, for not too small $n_h$. 
	In this way one obtains the direct link between the parameters of the dc conductivity (\ref{eq54}) 
and Eq.~(\ref{eq55}). \cite{Castro09}

The solid line in the inset of Fig.~3 illustrates $\sigma_{\alpha \alpha}^{\rm intra}({\bf q},\omega)$ 
at ${\bf q} \approx 0$ 
in a typical experimental situation in graphene, corresponding to the Fermi energy $E_{\rm F} = -0.5$ eV.
	However, to obtain good agreement with experiment in the infrared region, one must use Eq.~(\ref{eq52}),
together with Eq.~(\ref{eq62}) for the interband contribution.
	In such a phenomenological analysis, one starts with
the appropriate assumption for the imaginary part of the memory function $M^i_{1\alpha} (\omega)$ and then calculate
$M^r_{1\alpha} (\omega)$ by means of the Kramers--Kronig relations.
	The parameters in $M^i_{1\alpha} (\omega)$ obtained in this way are a clear indication that
the intraband relaxation-time approximation fails when the frequencies approach the infrared region.
	The comparison of the predictions of the relaxation-time approximation from Fig.~3 with  experimental data 
from Ref.~\onlinecite{Li08} at $\hbar \omega \approx E_{\rm F}$ leads to the same conclusion.

\begin{figure}[tb]
  \includegraphics[width=17pc]{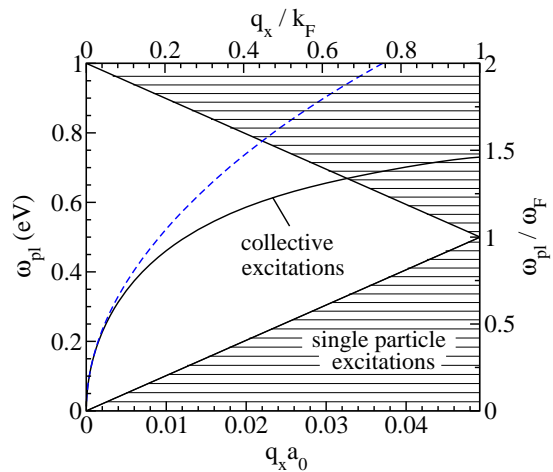}
  \caption{(Color online) Elementary excitations in hole-doped graphene in the random-phase approximation 
  [for $E_{\rm F}= \hbar \omega_{\rm F}= -0.5$ eV and ${\bf q} = (q_x,0)$].
  The solid and dashed lines represent the intraband plasmon dispersions calculated by using, respectively,
  $\varepsilon_\infty ({\bf q}, \omega)=1$ and $\tilde \varepsilon_\infty ({\bf q}, \omega)=1$ in Eqs.~(\ref{eq1})
  and (\ref{eq67}). 
  }
  \end{figure}

The inset of figure also shows how $\sigma_{\alpha \alpha}^{\rm intra}({\bf q}, \omega)$ from 
Eq.~(\ref{eq45}) depends on the wave vector ${\bf q}$ along the $\Gamma-K$ line in the first Brillouin zone.
	The intraband Landau damping is associated with the creation of real intraband
electron-hole pairs.
	The usual representation of the elementary excitations in hole-doped graphene in the 
ideal conductivity regime is shown in Fig.~4, including these excitations as well as
the real interband electron-hole pair excitations and the intraband plasmon modes.
	In both figures, $|\varepsilon_{L} ({\bf k}_+) - \varepsilon_{L} ({\bf k})| = q \hbar v_{\rm F}$ 
can be identified as the upper edge for the intraband electron-hole pair excitations
[$v_{\rm F} = (\sqrt{3} a t/2 \hbar)$  is the Fermi velocity].

\section{Transverse conductivity sum rule}
\subsection{Bare effective numbers of charge carriers}
The effective number $n_{\alpha \alpha}^{\rm intra,0}$ is shown in Fig.~5 as a function of the nominal concentration of conduction electrons $n$ 
and compared to the bare density of states 
\begin{eqnarray}
&& \hspace{-10mm}
\rho_0(E_{\rm F}) = \frac{1}{V} \sum_{L{\bf k} \sigma} \delta(\varepsilon_L ({\bf k})-E_{\rm F}).
\label{eq58}
\end{eqnarray}
	For the $\pi$ band almost empty, we obtain $n_{\alpha \alpha}^{\rm intra,0} \approx n$ 
[notice that $\gamma_{\alpha \alpha}^{\pi \pi} ({\bf k}) \approx 1$ and $m_{xx} \approx m$ in this case], 
the result which is typical of the ordinary 2D metallic systems.
	In this case, the dc conductivity $\sigma^{\rm dc }_{\alpha \alpha} = e n \mu$ is described indeed
in terms of the electron mobility $\mu$.
	On the other hand, in the Dirac regime $1.7 < V_0n \le 2$ [corresponding to 
$|{\bf v}^{\pi} ({\bf k})| \approx v_{\rm F}$], 
the result is $n_{\alpha \alpha}^{\rm intra,0} \approx (m|E_{\rm F}|/\hbar^2 \pi)$, or 
$V_0 n_{\alpha \alpha}^{\rm intra,0} \approx (m/m_{xx}) (3t/4) V_0 \rho_0(E_{\rm F})$, leading to 
$n_{\alpha \alpha}^{\rm intra,0} \propto \sqrt{n_h}$.
\cite{Castro09}

We can also calculate the bare total number of charge carriers $n^{\rm tot,0}_{\alpha \beta} ({\bf q})$
in two $2p_z$ bands by using the procedure from Sec.~IV,
\begin{eqnarray}
&& \hspace{-5mm} n^{\rm tot,0}_{\alpha \beta} ({\bf q}) = \sum_{LL'}
\frac{1}{V} \sum_{{\bf k} \sigma} \frac{m}{e^2}
J_\alpha^{LL'} ({\bf k},{\bf k}_+) 
J_{\beta}^{L'L} ({\bf k}_+,{\bf k}) 
\nonumber \\
&& \hspace{10mm} \times 
\frac{f_L({\bf k}) - f_{L'}({\bf k}_+)}{\varepsilon_{L'L} ({\bf k}_+,{\bf k})}.
\label{eq59}
\end{eqnarray}

These two effective numbers are expected to be of relevance in considering the electrodynamic properties 
of the doped graphene samples which are not too close to the ballistic conductivity regime.
	In the latter case, we have to use  the renormalized effective numbers 
$n^{\rm intra}_{\alpha \beta} ({\bf q})$ and $n^{\rm tot}_{\alpha \beta} ({\bf q})$, which are calculated by means of the renormalized Green's functions ${\cal G}_L ({\bf k}, {\it i} \omega_n )$. \cite{Peres08,KupcicUP}

\begin{figure}[tb]
  \includegraphics[width=18pc]{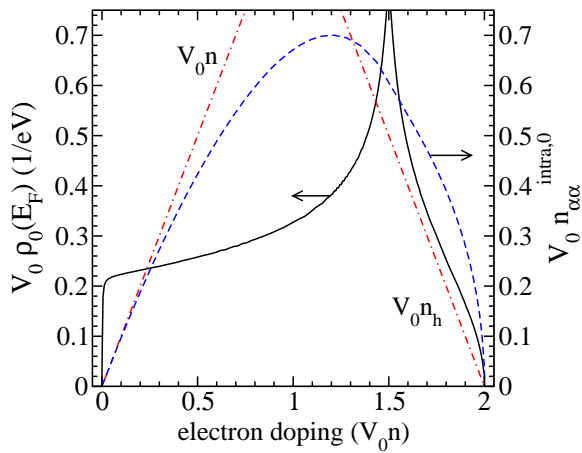}
  \caption{(Color online) The dependence of the effective number 
  $V_0 n^{\rm intra,0}_{\alpha \alpha}$, Eq.~(\ref{eq35}) with $n_L({\bf k})$ replaced by $f_L({\bf k})$, and 
  the density of states $V_0 \rho_0 (E_{\rm F})$, Eq.~(\ref{eq58}), on the electron doping in hole-doped graphene.
  Notice that $n^{\rm intra,0}_{\alpha \alpha} \approx n$ for $V_0 n \ll 2$ and that 
  $n^{\rm intra,0}_{\alpha \alpha} \approx (3t /4) \rho_0(E_{\rm F}) \propto \sqrt{n_h}$ for $V_0n_h \ll 2$.
  }
  \end{figure}

\subsection{Two-band dynamical conductivity}
In principle, the renormalized effective numbers $n^{\rm intra}_{\alpha \beta} ({\bf q})$ and 
$n^{\rm tot}_{\alpha \beta} ({\bf q})$ can be extracted from measured reflectivity data with the aid of
the transverse conductivity sum rule \cite{Pines89,Wooten72}
\begin{eqnarray}
&& \hspace{-5mm}
8 \int_0^\infty {\rm d} \omega \, \frac{1}{a_0}{\rm Re} \{ \sigma_{\alpha \alpha}^i ({\bf q}, \omega) \}
= V_0n^{i}_{\alpha \alpha} ({\bf q}) \Omega_0^2 = \big[ \Omega_{\rm pl}^i ({\bf q}) \big]^2,
\nonumber \\
\label{eq60}
\end{eqnarray}
$i = {\rm intra},  {\rm tot}$,
	Here, $\Omega_0 = \sqrt{4 \pi e^2/m a_0 V_0}$ is the auxiliary frequency scale.
	In the leading approximation, the transverse conductivity $\sigma_{\alpha \alpha}^{\rm tot} ({\bf q}, \omega)$ 
can be calculated by using Eq.~(\ref{eqA8}) in which the transverse current-dipole correlation function 
$\pi_{\alpha \tilde \alpha} ({\bf q}, \omega)$ is replaced by the longitudinal current-dipole correlation function 
$\pi_{\alpha \tilde \alpha} ({\bf q}, \omega) =({\it i}/q_\alpha) \pi_{\alpha 0} ({\bf q}, \omega)$.
\cite{Kubo95,Ziegler06,Kupcic13}

It must be emphasized that the sum rule (\ref{eq60}) is the general result, which is a direct consequence of
the Kubo formula (\ref{eq24}) [or the Ward identity relation (\ref{eq38})]
and the Kramers--Kronig relation for 
${\rm Re} \{ \sigma_{\alpha \alpha}^{\rm tot} ({\bf q}, \omega) \}$.
	The most important fact about this sum rule is that it is insensitive to details of
the scattering Hamiltonian $H'=H_1'+H_2'$, and, consequently, can be used as a simple direct test of
gauge invariance of the total conductivity formula.
	It is not hard to see that the expression (\ref{eq56}) for the ${\bf q} \approx 0$ intraband dielectric 
susceptibility is gauge invariant, at variance with the widely used expression (\ref{eq57}).

The semiphenomenological form of $\sigma_{\alpha \alpha}^{\rm tot} ({\bf q}, \omega)$ which is consistent 
with this general result is
\begin{eqnarray}
&& \hspace{-10mm} 
\sigma_{\alpha \alpha}^{\rm tot}({\bf q}, \omega)= \sigma_{\alpha \alpha}^{\rm intra} ({\bf q}, \omega)
+\sigma_{\alpha \alpha}^{\rm inter} ({\bf q}, \omega),
\label{eq61}
\end{eqnarray}
where $\sigma_{\alpha \alpha}^{\rm intra} ({\bf q},\omega)\approx \sigma_{\alpha \alpha}^{\rm intra} (\omega)$ 
and $\sigma_{\alpha \alpha}^{\rm inter} ({\bf q},\omega)\approx \sigma_{\alpha \alpha}^{\rm inter} (\omega)$ 
are given, respectively,  by Eq.~(\ref{eq53}) and by
\begin{eqnarray}
&& \hspace{-10mm} 
\sigma_{\alpha \alpha}^{\rm inter} ({\bf q}, \omega) =
\frac{1}{V} \sum_{L\neq L'} \sum_{{\bf k} \sigma}
\frac{{\it i} \hbar |J^{LL'}_{\alpha}({\bf k},{\bf k}_+)|^2}{\varepsilon_{L'L}({\bf k}_+,{\bf k})}
\nonumber \\
&& \hspace{10mm} \times 
\frac{n_L({\bf k})-n_{L'}({\bf k}_+)}{\hbar \omega + \varepsilon_{LL'}({\bf k},{\bf k}_+) + {\it i}\hbar \Gamma_2}.
\label{eq62}
\end{eqnarray}
	The total (two-band) conductivity obtained in this way is illustrated in Fig.~3 by the solid and 
the dot-dot-dashed line.
	It is worth noticing that, although the experimental relation $\sigma^{\rm dc}_{\alpha \alpha} \propto n_h$ 
suggests that the number $n_h$ is the effective number of charge carriers that participate in the low-energy physics of hole-doped graphene, the intraband transverse conductivity sum rule shows that this effective number is actually equal to 
$n^{\rm intra}_{\alpha \alpha}$.

The interband memory function can be introduced phenomenologically by replacing the damping energy 
${\it i} \Gamma_2$ in Eq.~(\ref{eq62}) with $M^{LL'} ({\bf k},{\bf q},\omega)$,  $L \neq L'$.
	In the simplest approximation,
it is the sum of the electron self-energy from the upper band and the hole self-energy from the lower band. 
	Although the corresponding vertex corrections are important (for example, in explaining the occurrence 
of the Wannier excitons in a general case), they are usually neglected.
	In graphene, this simplification is incorrect, and, consequently, requires reconsiderations because 
the energy difference $\varepsilon_{LL'}({\bf k},{\bf k}_+)$ in the denominator of Eq.~(\ref{eq62}) becomes very small 
for $|E_{\rm F}| \rightarrow 0$, leading to $\sigma_{\alpha \alpha}^{\rm inter}({\bf q}, \omega \approx 0) \neq 0$ 
in this limit. \cite{Ziegler06,Lewkowicz09}

\section{Energy loss function}
\subsection{Plasma oscillations}
It is apparent that in simple two-band models the extended generalized Drude formula (\ref{eq61})
can support two different low-frequency collective modes. \cite{Platzman73}
	The first one, usually called the intraband plasmon (the Dirac plasmon in graphene), involves the oscillations 
of doped holes/electrons, with the frequency $\omega_{\rm pl} ({\bf q})$ proportional to the square root 
of the effective number $n_{\alpha \alpha}^{\rm intra} ({\bf q},\omega_{\rm pl} ({\bf q}))$.
	In the leading approximation, this effective number is obtained by expanding Eq.~(\ref{eq45}) 
in powers of $q_\alpha$ and writing the result in the form
\begin{eqnarray}
&& \hspace{-5mm} 
\sigma_{\alpha \alpha}^{\rm intra} ({\bf q}, \omega) \approx
\frac{{\it i} e^2 }{\omega + M_{1\alpha}(\omega)} \frac{1}{V} \sum_{L{\bf k} \sigma}
 [v_\alpha^L ({\bf k})]^2 \bigg(-\frac{\partial n_L({\bf k})}{\partial \varepsilon_L ({\bf k})}\bigg)
\nonumber \\
&& \hspace{15mm} \times 
\bigg[1 + \bigg(\frac{v_\alpha^L({\bf k}) q_\alpha}{\omega} \bigg)^2 + \ldots \bigg]
\nonumber \\
&& \hspace{13mm}
= \frac{{\it i} e^2 }{m(\omega + M_{1\alpha}(\omega))}n^{\rm intra}_{\alpha \alpha}({\bf q}, \omega).
\label{eq63}
\end{eqnarray}
	At a crude level, $n^{\rm intra}_{\alpha \alpha}({\bf q}, \omega) \approx n^{\rm intra}_{\alpha \alpha}[1 
+ 3 \langle [v_\alpha^L({\bf k})]^2 \rangle q^2 / \omega^2]$ can be approximated by $n^{\rm intra}_{\alpha \alpha}$
from Eq.~(\ref{eq35}).

On the other hand, in the second mode all electrons from the two  bands oscillate
with a much higher frequency $\omega_{\rm pl}^{\rm tot} ({\bf q})$, which is proportional to 
$\sqrt{n_{\alpha \alpha}^{\rm tot}({\bf q},\omega_{\rm pl}^{\rm tot} ({\bf q}))}$.
	The effective number $n_{\alpha \alpha}^{\rm tot}({\bf q},\omega_{\rm pl}^{\rm tot} ({\bf q}))$ 
is obtained by expanding Eq.~(\ref{eq61}) in powers of $q_\alpha$.
	$n_{\alpha \alpha}^{\rm tot}({\bf q})$ from Eq.~(\ref{eq59}), taken at ${\bf q}=0$, 
is the leading contribution to this number.

Strictly speaking, these two plasmon frequencies correspond to two roots of the longitudinal dielectric function 
${\rm Re} \{ \varepsilon ({\bf q}, \omega) \}$.
	In multiband electronic systems, the frequency of the intraband plasmon is finite only if
one of the bands is partially full.
	It is also evident that the second plasmon is clearly visible in
${\rm Re} \{ \varepsilon ({\bf q}, \omega) \}$ only if the bands in question are narrow and the direct interband threshold energy is not too high. \cite{Landau95}
	Only in this case the ''interband'' plasmons cannot decay directly into interband electron-hole pair excitations.

For frequencies $\omega \approx \omega_{\rm pl}({\bf q})$, the inverse of the dielectric function of graphene 
and the screened long-range interaction $\tilde v({\bf q}, \omega)$ can be shown in the form
\cite{Hwang07,Despoja13}
\begin{eqnarray}
&& \hspace{-10mm}
\frac{1}{\varepsilon ({\bf q}, \omega)} = \frac{ \omega^2/\varepsilon_\infty}{
\omega^2-\omega_{\rm pl}^2({\bf q}, \omega) + {\it i} \omega \Gamma_{\rm pl} ({\bf q}, \omega)},
\nonumber \\
&& \hspace{-10mm}
\tilde v({\bf q}, \omega) = \frac{v({\bf q})}{\varepsilon ({\bf q}, \omega)} 
= \frac{ (\omega^2 /\varepsilon_\infty)v({\bf q})}{
\omega^2-\omega_{\rm pl}^2({\bf q}, \omega) + {\it i} \omega \Gamma_{\rm pl} ({\bf q}, \omega)}.
\label{eq64}
\end{eqnarray}
	The Dirac plasmon frequency $\omega_{\rm pl} ({\bf q})$ is a root of ${\rm Re} \{ \varepsilon ({\bf q}, \omega) \}$.
	It comprises three contributions,
\begin{eqnarray}
&& \hspace{-5mm}
\omega_{\rm pl}^2 ({\bf q}) \approx  [\omega_{\rm pl}^0 ({\bf q})]^2
+ \frac{2\pi q}{\varepsilon_\infty}\omega_{\rm pl} ({\bf q}) 
{\rm Im}\{ \sigma_{\alpha \alpha}^{\rm inter} ({\bf q},\omega_{\rm pl}({\bf q})) \}
\nonumber \\
&& \hspace{5mm}
+ \frac{2\pi q}{\varepsilon_\infty} \big[ {\rm Re}\{ \pi_{\alpha \alpha}^{{\rm intra}} ({\bf q},\omega_{\rm pl}({\bf q}))\} 
- \pi_{\alpha \alpha}^{{\rm intra}} ({\bf q}) \big].
\label{eq65}
\end{eqnarray}
	The first one is the square of the bare frequency
$[\omega_{\rm pl}^0 ({\bf q})]^2 = (2\pi e^2 q/m \varepsilon_\infty) 
n^{\rm intra}_{\alpha \alpha}({\bf q})$, with small ${\bf q}$ dependent corrections included
[notice that the model for the dc conductivity (\ref{eq55}) is consistent with the relation
$[\omega_{\rm pl} ({\bf q})]^2 \approx  (2\pi e^2 q/m) n_h$].
	The second one describes the dynamical screening effects and the third one presumably small residual terms.
	Any complete treatment of the Dirac plasmons should include the estimation of this residual contribution.

On the other hand, the damping effects come from the direct and indirect intraband and interband absorption
processes in
\begin{eqnarray}
&& \hspace{-10mm}
\hbar \Gamma_{\rm pl} ({\bf q}, \omega)= q a_0 \frac{2 \pi \hbar}{a_0 \varepsilon_\infty} 
{\rm Re} \{\sigma_{\alpha \alpha}^{\rm tot} ({\bf q},\omega) \}.
\label{eq66}
\end{eqnarray}
	As mentioned above, the relaxation-time approximation gives a reasonable description of the direct 
absorption processes, but underestimates the indirect absorption processes typically one order of magnitude.
	Therefore, the detailed study of the damping energy $\hbar \Gamma_{\rm pl} ({\bf q}, \omega)$
requires the theory beyond the relaxation-time approximation, the one which is capable of explaining 
both the $\omega=0$ part in $M^i_{1\alpha}(\omega)$, $M^i_{1\alpha}(0)$, and the frequency dependent corrections 
$\Delta M^i_{1\alpha}(\omega)$ [$M^i_{1\alpha}(\omega) = M^i_{1\alpha}(0)+\Delta M^i_{1\alpha}(\omega)$].
	Nevertheless, a good quantitative understanding of the energy loss measurements is possible by inserting
${\rm Re} \{\sigma_{\alpha \alpha}^{\rm tot} ({\bf q},\omega) \}$ [or $M^i_{1\alpha}(\omega)$], 
taken from reflectivity measurements, into Eq.~(\ref{eq66}). 

\begin{figure}[tb]
  \includegraphics[width=15pc]{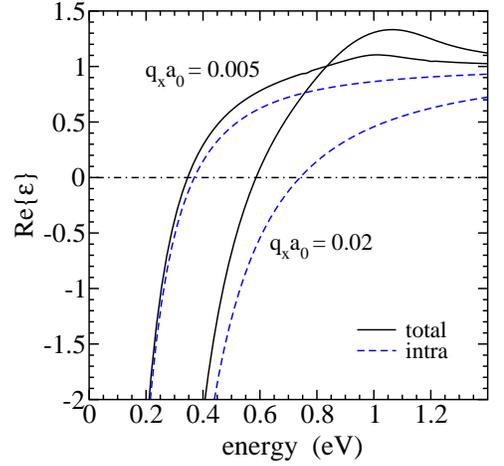}
  \caption{(Color online) The dependence of the real part of the dielectric function (\ref{eq67}) on $\hbar \omega$, 
   for $E_{\rm F} = -0.5$ eV and for two values of the wave vector ${\bf q} = (q_x,0)$, $q_x a_0 = 0.005$ and 0.02.
   The solid and dashed lines correspond, respectively, to $\varepsilon_\infty ({\bf q}, \omega)=1$ and 
   $\tilde \varepsilon_\infty ({\bf q}, \omega)=1$.
  }
  \end{figure}

For simplicity we rewrite $\varepsilon ({\bf q}, \omega)$ from Eq.~(\ref{eq1}) in the form
\begin{eqnarray}
&& \hspace{-10mm}
\varepsilon ({\bf q}, \omega) \approx  \tilde \varepsilon_\infty ({\bf q}, \omega) 
+ v({\bf q}) \sum_{\alpha}  \frac{{\it i}}{\omega} q_\alpha^2 \sigma_{\alpha \alpha}^{\rm intra} ({\bf q}, \omega),
\label{eq67}
\end{eqnarray}
where $\sigma_{\alpha \alpha}^{\rm intra} ({\bf q}, \omega)$ is given by Eq.~(\ref{eq45}), 
and in the numerical calculations we use the relaxation-time approximation.
	Figure~6 illustrates the real part of $\varepsilon ({\bf q}, \omega)$ in hole-doped graphene for 
$E_{\rm F} = -0.5$ eV, $q_x a_0 =0.01$ and 0.03, and $\varepsilon_\infty=1$.
	As mentioned above, to estimate $\omega_{\rm pl} ({\bf q})$  independently, we multiply the frequency 
$\Omega_{\rm pl}^{\rm intra}({\bf q}=0) = \sqrt{(e^2/2a_0) 8|E_{\rm F}|}/\hbar$ 
from Eq.~(\ref{eq60}) by $\sqrt{q a_0/2} \sqrt{1 +(3/2)(v_{\rm F} q/\omega)^2}$.
	For $\hbar \omega_{\rm pl} ({\bf q}) < |E_{\rm F}|$, the agreement between
this frequency (dashed lines in the figure) and the result of the former approach (solid lines) is surprisingly good considering
that the real and imaginary part of Eq.~(\ref{eq67}) are both complicated functions of $\omega$ and ${\bf q}$.
\cite{Despoja13}
	The dominant correction to $\omega_{\rm pl}^0 ({\bf q}) \approx \sqrt{q a_0/2} \, \Omega_{\rm pl}^{\rm intra}$
comes from the dynamical screening effect.
	This effect, together with the interband Landau damping, is also responsible for the disappearance of the second 
(''interband'') plasmon mode in graphene.

\begin{figure}[tb]
  \includegraphics[width=17pc]{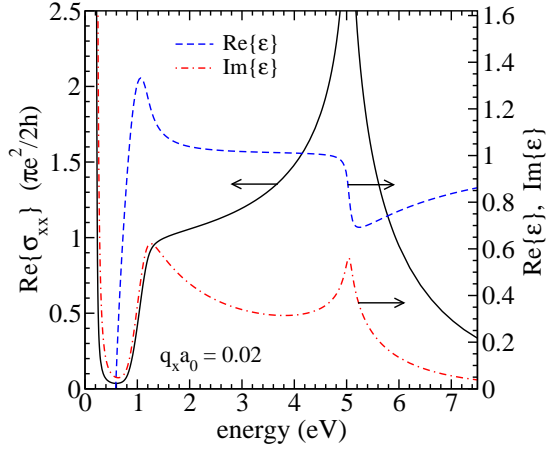}
  \caption{(Color online) Solid line: the real part of the dynamical conductivity, Eq.~(\ref{eq61}), 
  in hole-doped graphene, for $E_{\rm F} = -0.5$ eV, $\hbar \Gamma_1 = 10$ meV, $\hbar \Gamma_2 = 50$ meV, 
  $T=150$ K, and $q_xa_0=0.02$.
  Dashed and dot-dashed lines: the corresponding real and imaginary part of the dielectric function.
  }
  \end{figure}

\subsection{Dirac and $\pi$ plasmons}
The energy loss function $-{\rm Im} \{1/\varepsilon({\bf q}, \omega) \}$ 
is primarily useful for studying collective modes of electronic subsystem.
	Figure 7 illustrates ${\rm Re} \{\sigma_{xx}({\bf q}, \omega) \}$ and the corresponding functions 
${\rm Re} \{\varepsilon({\bf q}, \omega) \}$ and 
${\rm Im} \{\varepsilon({\bf q}, \omega) \}$ for $E_{\rm F} = -0.5$ eV in the 0$-$7.5 eV energy range.
	This figure shows that the van Hove singularity in the density of states $\rho (\mu)$ at 
$\hbar \omega = - E_{\rm vH} \approx 2.5$ eV is accompanied by the singularity in both
${\rm Re} \{\sigma_{\alpha \alpha}({\bf q}, \omega) \}$ and ${\rm Im} \{\varepsilon({\bf q}, \omega) \}$ 
at $\hbar \omega \approx 5$ eV and by the sharp decrease in ${\rm Re} \{\varepsilon({\bf q}, \omega) \}$ 
in the same energy region.
	The resulting function $-{\rm Im} \{1/\varepsilon({\bf q}, \omega) \}$ is shown in Figs.~8 and 9.
	There are two distinctly resolved maxima in this function.
	The first one is placed at the Dirac plasmon energy $\hbar \omega \approx \hbar \omega_{\rm pl} ({\bf q})$
and the second one at $\hbar \omega \approx -2E_{\rm vH} \approx 5$ eV.
	Therefore, the first maximum is related to the first zero of 
${\rm Re} \{\varepsilon({\bf q}, \omega) \}$ and illustrates the frequency and the damping energy 
of the Dirac plasmon from Eqs.~(\ref{eq65}) and (\ref{eq66}).
	On the other hand, the second maximum (usually called the $\pi$ plasmon) is simply a consequence 
of the singularity in the single-electron density of states.
	Its position and half-width are both complicated functions of the parameters in 
$\sigma_{xx}^{\rm tot}({\bf q}, \omega)$.
	Evidently the latter maximum is absent in the Dirac cone approximation.

\begin{figure}[tb]
  \includegraphics[width=15pc]{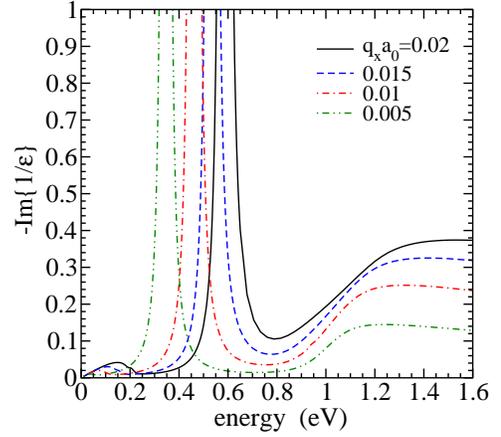}
  \caption{(Color online) The Dirac plasmon peak in the energy loss function for different values of $q_xa_0$.
  The parameters are the same as in Fig.~7.
  }
  \end{figure}

\begin{figure}[tb]
  \includegraphics[width=15pc]{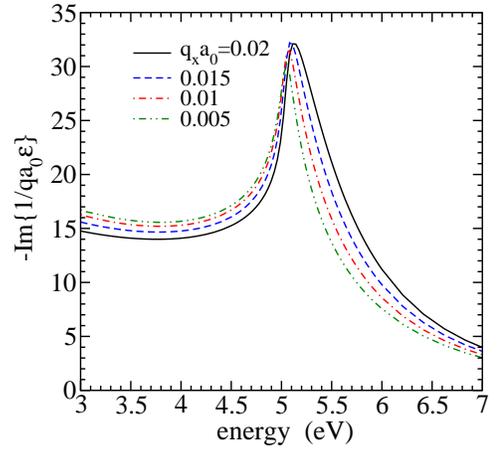}
  \caption{(Color online) The $\pi$ plasmon peak in the energy loss function for different values of ${\bf q}=(q_x,0)$.
  The spectra are divided by $q a_0$ for clarity.
  The imaginary part of the screened long-range Coulomb interaction is 
  ${\rm Im}\{ \tilde v({\bf q}, \omega)\} = 2\pi a_0 {\rm Im}\{1/q a_0 \varepsilon ({\bf q}, \omega)\}$.
  }
  \end{figure}

\subsection{Microscopic treatment of relaxation processes}
The comparison of Fig.~9 with the experimental data from Ref.~\onlinecite{Eberlein08}
shows that the relaxation-time approximation can be safely used
in describing the $\pi$ plasmon structure in the energy loss function.
	On the other hand, it gives only an oversimplified description of the damping of Dirac plasmons,
as already mentioned.
	Nevertheless, for $\hbar \omega \approx \hbar \omega_{\rm pl}({\bf q}) < |E_{\rm F}|$, we can treat
the damping energy $\hbar \Gamma_1$ as a fitting parameter.
	For example, $\hbar \Gamma_1 =0.02$ eV [which is a factor of 2.5 larger than
$\hbar M^{i}_{1\alpha}(0)$ extracted from the dc resistivity] corresponds to the typical experimental value
${\rm Re} \{\sigma_{\alpha \alpha} (0.2\,{\rm eV}) \} \approx 0.2 \, (\pi e^2/2 h)$. \cite{Li08}
	The inset of Fig.~10 illustrates the energy loss function 
$-{\rm Im} \{ 1/\varepsilon ({\bf q}, \omega) \}$ obtained in this way for $E_{\rm F} = -0.4$ eV.
	The result for $qa_0 = 0.001$ and $0.0015$ is in reasonably good agreement with experiment. \cite{Yan13}

\begin{figure}
   \centerline{\includegraphics[width=20pc]{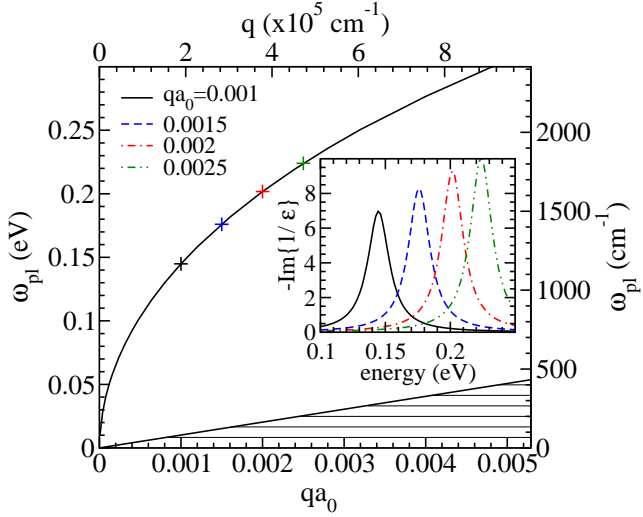}}
   \caption{(Color online) Main figure: the dispersion of the Dirac plasmons for $E_{\rm F}=-0.4$ eV.
   Inset of figure: the energy loss function $-{\rm Im} \{ 1/\varepsilon({\bf q}, \omega \}$ calculated by means 
   of the relaxation-time approximation, for $\hbar \omega$ close to the energy of the in-plane optical phonons 
   and for $\hbar \Gamma_1 = 0.02$ eV.
   }
  \end{figure}
  
An alternative to this oversimplified description of the damping effects at  $\hbar \omega < |E_{\rm F}|$ is the microscopic
memory-function approach. \cite{Gotze72,Kupcic13,Kupcic14}
	In this approach the intraband memory function is calculated by using the high-energy expansion of the 
RPA irreducible $4 \times 4$ current-current correlation functions $\pi_{\mu \nu}^{\rm intra} ({\bf q}, \omega)$
in Eq.~(\ref{eqA7}).
	The contributions to the correlation functions $\pi_{\mu \nu}^{\rm intra} ({\bf q}, \omega)$
which are second order in $H'$ are shown for the boson-mediated electron-electron interactions and for the non-retarded electron-electron interactions in Figs.~11 and 12, respectively.
	In the intraband scattering approximation for the short-range electron-electron interactions, 
the explicit calculation of these two contributions to $\pi_{\alpha \alpha}^{\rm intra} ({\bf q}, \omega)$ leads to 
the intraband memory function 
$M_{\alpha}^{LL} ({\bf k},\omega) \approx M^{[2]}_{1\alpha} ({\bf k}, \omega)+ M^{[4]}_{1\alpha} ({\bf k},\omega) 
+ \Delta M_{\alpha}^{LL} ({\bf k},\omega)$, where
\begin{eqnarray}
&& \hspace{-5mm}
\hbar M^{[2]}_{1\alpha} ({\bf k}, \omega) = -\frac{1}{N} \sum_{L'\nu {\bf k'}} 
|G_\nu^{LL'}({\bf k},{\bf k}')|^2 \bigg(1 -\frac{v_\alpha^{L'} ({\bf k}')}{v_\alpha^L ({\bf k})} \bigg) 
\nonumber \\
&& \hspace{0mm} \times 
\sum_{s = \pm 1} \sum_{s' = \pm 1} \frac{s'\big[f^b (s'\omega_{\nu {\bf k}-{\bf k}'}) 
+ f(s \varepsilon_{L'}({\bf k}')) \big]}{
\hbar \omega + {\it i} \eta + s \varepsilon_{LL'}({\bf k},{\bf k}') 
+ s' \hbar \omega_{\nu {\bf k}-{\bf k}'}},
\label{eq68}
\\
&& \hspace{-5mm}
\hbar M^{[4]}_{1\alpha} ({\bf k},\omega) = 
-\sum_{{\bf k'}{\bf q} \sigma'} \frac{|\varphi_{\sigma \sigma'}({\bf q})|^2}{V^2} 
 \frac{1}{v_\alpha^L ({\bf k})}\big[v_\alpha^L ({\bf k}) + v_\alpha^L ({\bf k}'_+)
\nonumber \\
&& \hspace{5mm}
- v_\alpha^L ({\bf k}') -v_\alpha^L ({\bf k}_+)  \big][f(\varepsilon_L({\bf k}'))-f(\varepsilon_L({\bf k}'_+))]
\nonumber \\
&& \hspace{0mm}
\times \sum_{s = \pm 1} \frac{f^{b}(\omega({\bf k}'_+,{\bf k}')) + f(\varepsilon_L({\bf k}_+))}{
\hbar \omega + {\it i} \eta + s \varepsilon_{LL}({\bf k},{\bf k}') + s \varepsilon_{LL}({\bf k}'_+,{\bf k}_+)},
\label{eq69}
\end{eqnarray}
with $\omega_{LL}({\bf k}'_+,{\bf k}')=\varepsilon_{LL}({\bf k}'_+,{\bf k}')/\hbar$.

The plasmon damping rate (\ref{eq66}) describes in the first place
the decay of the plasmons into electron-hole excitations.
	For example, in the process of the decay of the Dirac plasmons from Fig.~10 
an electron goes from a filled state ${\bf k}$ to an empty state ${\bf k}'$ with conservation of energy and momentum.
	These processes are usually called the indirect absorption processes.
	According to Eqs.~(\ref{eq68}) and (\ref{eq69}), they describe the creation of one electron-hole pair 
in combination with another elementary excitation (acoustic or optical phonon, or second electron-hole pair).
	Although these processes are missing in the RPA-like illustration in Fig.~4, they
play an essential role in the microscopic explanation of the damping effects in the region which is far away from the Landau damping.

\begin{figure}[tb]
   \centerline{\includegraphics[width=18pc]{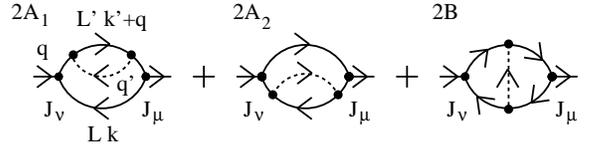}}
   \caption{Three $(H_1')^2$ contributions to $\pi_{\mu \nu}^{\rm intra} ({\bf q}, \omega)$, labeled 
   by $2A_1$ (electron self-energy term), $2A_2$ (hole self-energy term), and $2B=2B_1+2B_2$ (vertex correction).
   }
   \end{figure}

\begin{figure}[tb]
   \centerline{\includegraphics[width=20pc]{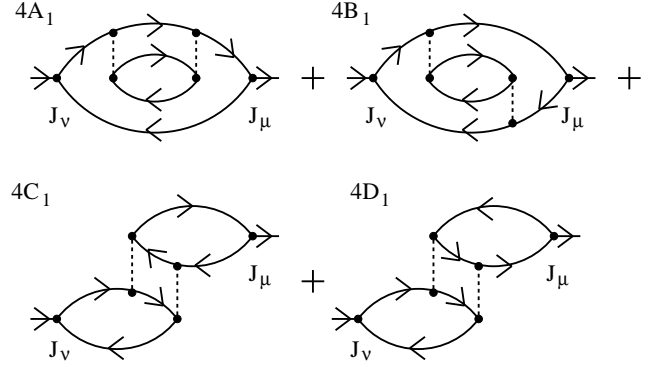}}
   \caption{Four contributions to $\pi_{\mu \nu}^{\rm intra} ({\bf q}, \omega)$ 
   out of eight contributions that are proportional to $(H_2')^2$ [or $(H_1')^4$].
   }
   \end{figure}

We defer a full discussion of the microscopic memory-function approach to Ref.~\onlinecite{KupcicUP}.
	Here we only underline the most important qualitative conclusions.
	(i) The vertex corrections (diagrams $2B$ and $4B$) are responsible for the exact cancellation of all 
retarded and non-retarded (${\bf q} \approx 0$) forward scattering contributions to 
$M^{LL}_\alpha ({\bf k},\omega)$.
	(ii) This conclusion holds for the scattering by intraband plasmons as well, and makes the analysis of 
$M^{LL}_\alpha ({\bf k},\omega)$ much simpler than the analysis of the corresponding single-electron self-energy 
$\Sigma_{L} ({\bf k},\omega)$.
	(iii) The Aslamazov--Larkin contributions (diagrams $4C$ and $4D$) 
\cite{Vollhardt80,Bergeron11} lead to the strong suppression 
of the normal backward scattering processes, and, therefore, causes a further reduction in
$M^{LL}_\alpha ({\bf k},\omega)$.
	In simple weakly interacting systems the result is the imaginary part of $M^{LL}_\alpha ({\bf k},\omega)$ 
which is dominated by the Umklapp backward scattering processes, in agreement with the common Fermi liquid theory.
\cite{Ziman72,Pines89,Abrikosov88}
	(iv) The situation is distinctly different for hole-doped graphene because the intensity 
of the Umklapp scattering processes is in general very sensitive to the size and 
the shape of the Fermi surface.

The comparison with the results of the energy loss measurements at energies $\hbar \omega$ comparable 
to the energy of the in-plane optical phonons $\hbar \omega_{\nu {\bf q}}$ (the case illustrated 
in the inset of Fig.~10) shows that the scattering by disorder, by acoustic phonons, and by other electrons
can be represented by $\hbar \Gamma_1^0$, which is nearly independent of frequency 
($\approx \hbar \Gamma_1 = 0.02$ eV in Fig.~10).
	On the other hand, the scattering by optical phonons in $M^{[2]}_{1 \alpha} ({\bf k},\omega)$ produces 
strong frequency dependent effects for 
$\hbar \omega_{\rm pl} ({\bf q}) \approx \hbar \omega_{\nu {\bf q}} < |E_{\rm F}|$, and, therefore,
this scattering channel requires the detailed numerical analysis.
	In order to understand the role of the vertex corrections in this scattering channel in detail, 
we have to explain quantitatively not only the frequency dependence of $M^{LL}_{1 \alpha} ({\bf k},\omega)$ but also the frequency dependence of the single-electron self-energy  $\Sigma_{L} ({\bf k},\omega)$ extracted from ARPES measurements \cite{Bostwick07,Pletikosic12}.
	This question is left for future studies.

\section{Conclusion}
The Ward identity relation has been proven here for a general multiband electronic model using 
the zero temperature formalism.
	It is shown that this relation leads to the same relations among the elements of the four-current 
response tensor as the first and the second Kubo formula for the conductivity tensor.
	The general criteria for occurrence of the intraband and ''interband'' plasmon modes are briefly discussed as well.

We apply then the results to hole-doped graphene, and determine the dispersions and the damping parameters for
the long-wavelength Dirac and $\pi$ plasmons.
	We have demonstrated that it is possible to explain consistently the damping of these collective modes, 
the relaxation processes in the dynamical conductivity, and the single-electron self-energy in ARPES spectra, 
even within the relaxation-time approximation.
	It is pointed out that the single-electron propagators are strongly affected by the forward scattering 
processes, in particular by the scattering by two-dimensional intraband plasmon modes.
	On the other hand, these scattering processes are cancelled identically in any gauge invariant form 
of the intraband conductivity tensor.

The semiphenomenological memory-function conductivity model, when treated consistently 
with the general Ward identity, is able to capture 
all aspects of the retarded and non-retarded electron-electron interactions in weakly interacting electronic systems.
	To extend the theory to systems with strong local and/or short-range interactions
we must use a more accurate treatment of the intraband and interband electron-hole propagators.
	We shall give in the accompanying article \cite{KupcicUP}
both the detailed description of the response theory beyond the relaxation-time approximation and the quantitative analysis of the memory function in hole-doped graphene.

\section*{Acknowledgment}
Source files of published data provided by V. Despoja are greatfully acknowledged.
This research was supported by the Croatian Ministry of Science 
and Technology under Project 119-1191458-0512.

\appendix
\section{Kubo formulae}
Electrodynamic properties of a general electronic system with multiple bands at the Fermi level 
are naturally described in terms of two real-time density correlation functions \cite{Kubo95}
\begin{eqnarray}
&& \hspace{-7mm}
\widetilde \chi({\bf q},\omega) = \int_0^\infty {\rm d} t\, {\rm e}^{{\it i} \omega t} \frac{1}{V} 
\frac{1}{{\it i} \hbar} \big<  \big[ \hat \rho ({\bf q},t), \hat \rho (-{\bf q},0) \big]  \big>,
\label{eqA1}
\\
&& \hspace{-10mm}
\widetilde \sigma_{\alpha \beta} ({\bf q}, \omega) = 
\beta \int_0^\infty {\rm d} t\, {\rm e}^{{\it i} \omega t} \frac{1}{V} 
\big< \hat J_\beta (-{\bf q},0); \hat J_\alpha ({\bf q},t) \big> .
\label{eqA2}
\end{eqnarray}
	The former one is the screened dielectric susceptibility and the latter one is the 
screened dynamical conductivity tensor.
	The relations (\ref{eqA1}) and (\ref{eqA2}) are also known as the Kubo formula for dielectric
susceptibility and the Kubo formula for conductivity, respectively.
	The susceptibility $\chi({\bf q},\omega)$ and the conductivity tensor
$\sigma_{\alpha \beta} ({\bf q}, \omega)$ are simply the RPA irreducible parts of 
$\widetilde \chi({\bf q},\omega)$ and
$\widetilde \sigma_{\alpha \beta} ({\bf q}, \omega)$.
\cite{Kubo95}

It is also useful to introduce the notation 
$\hat J_{\tilde \alpha} ({\bf q}) = - \hat P_{\alpha} ({\bf q})$, where $ \hat P_{\alpha} ({\bf q})$ is the dipole density operator and 
$J^{LL'}_{\tilde \alpha} ({\bf k}, {\bf k}_+) = -P^{LL'}_{\alpha} ({\bf k}, {\bf k}_+)$ is the related vertex function.
	$P^{LL}_\alpha ({\bf k}_+,{\bf k}) = - P^{LL}_\alpha ({\bf k},{\bf k}_+) \equiv p_\alpha({\bf q}) 
= -{\it i}e/q_\alpha$ is the intraband dipole vertex function.
	It is not hard to show that for an arbitrary orientation of the wave vector ${\bf q}$,
${\bf q} = \sum_\alpha q_\alpha \hat e_\alpha$, the dipole vertex function is connected to that from Eq.~(\ref{eq16}) by the relations
\cite{Kupcic13}
\begin{eqnarray}
&& \hspace{-10mm}
e q^{LL'} ({\bf k}, {\bf k}_+) = -{\it i} \sum_\alpha q_\alpha P_\alpha^{LL'} ({\bf k},{\bf k}_+) 
\nonumber \\
&& \hspace{10mm}
= \sum_\alpha q_\alpha \frac{ \hbar J_\alpha^{LL'} ({\bf k},{\bf k}_+)}{\varepsilon_{L'L} ({\bf k}_+,{\bf k})}.
\label{eqA3}
\end{eqnarray}
	Notice that this relation can also be shown in the form
\begin{eqnarray}
&& \hspace{-10mm}
\sum_\alpha q_\alpha J_\alpha^{LL'} ({\bf k},{\bf k}_+) - \omega e q^{LL'} ({\bf k}, {\bf k}_+) 
\nonumber \\
&& \hspace{0mm}
= [- \omega + \varepsilon_{L'L} ({\bf k}_+,{\bf k}) / \hbar  ] e q^{LL'} ({\bf k}, {\bf k}_+),
\label{eqA4}
\end{eqnarray}
with $\sum_\alpha q_\alpha P_\alpha^{LL'} ({\bf k},{\bf k}_+)  = {\it i} e q^{LL'} ({\bf k}, {\bf k}_+)$.

The definitions (\ref{eqA1}) and (\ref{eqA2}), together with the two basic relations from macroscopic electrodynamics
\cite{Landau95}
\begin{eqnarray}
&& \hspace{-10mm}
{\bf E}({\bf r}, t) = - \frac{\partial V^{\rm tot} ({\bf r}, t) }{\partial {\bf r}}
-\frac{1}{c} \frac{\partial {\bf A}^{\rm tot}({\bf r}, t)}{\partial t},
\label{eqA5}
\\
&& \hspace{-10mm}
\nabla \cdot {\bf J} ({\bf r}, t) 
+ \frac{\partial \rho ({\bf r}, t)}{\partial t} = 0,
\label{eqA6}
\end{eqnarray}
lead now to \cite{Kubo95}
\begin{eqnarray}
&& \hspace{-10mm}
\chi ({\bf q}, \omega) \equiv \pi_{00} ({\bf q}, \omega) 
= \frac{1}{{\it i}\omega} \sum_{\alpha \beta} q_\alpha \sigma_{\alpha \beta} ({\bf q}, \omega) q_\beta
\nonumber \\
&& \hspace{2mm}
= \frac{1}{\omega} \sum_{\alpha} q_\alpha \pi_{\alpha 0} ({\bf q}, \omega)
\nonumber \\
&& \hspace{2mm}
= \frac{1}{\omega^2} \sum_{\alpha \beta} 
q_\alpha \big[ \pi_{\alpha \beta} ({\bf q}, \omega)-\pi_{\alpha \beta} ({\bf q})\big] q_\beta,
\label{eqA7}
\\
&& \hspace{-13mm}
\sigma_{\alpha \beta} ({\bf q}, \omega) =  \pi_{\alpha \tilde \beta} ({\bf q}, \omega).
\label{eqA8}
\end{eqnarray}
	The expression (\ref{eqA5}) represents the gauge-invariant form of the macroscopic electric field 
${\bf E}({\bf r}, t)$, $V^{\rm tot} ({\bf r}, t)$ and ${\bf A}^{\rm tot}({\bf r}, t)$ are the screened scalar and vector potentials, 
and Eq.~(\ref{eqA6}) is the charge continuity equation.
	Equations (\ref{eqA7}) and (\ref{eqA8}) are the Kubo expressions for the RPA irreducible response
functions $\chi ({\bf q}, \omega)$ and $\sigma_{\alpha \beta} ({\bf q}, \omega)$.
	The same expressions are derived in the main text by integration by parts of the Fourier transform 
of the monopole-monopole correlations function $\pi_{00} ({\bf q}, t)$.

\section{Minimal substitution}
The second way of obtaining the relations (\ref{eqA7}) and (\ref{eqA8}) is to calculate 
the current $J_\mu ({\bf r}, t)$ induced in the medium by the vector and scalar potentials 
${\bf A}^{\rm tot}({\bf r}, t)$ and  $V^{\rm tot} ({\bf r}, t)$.
	The coupling of these fields to the electronic subsystem is described by the
coupling Hamiltonian $H^{\rm ext} = H^{\rm ext}_1+H^{\rm ext}_2 $ obtained by means 
of the gauge-invariant minimal substitution, where
\cite{Schrieffer64,Kupcic07}
\begin{eqnarray}
 && \hspace{-10mm}
H^{\rm ext}_1  =  -\frac{1}{c} \sum_{{\bf q} \mu} A_{\mu} ({\bf q})
\hat J_{\mu} (-{\bf q}),
\nonumber \\
&& \hspace{-10mm}
H^{\rm ext}_2 =  \frac{e^2}{2mc^2} \sum_{{\bf q} {\bf q}'\alpha \beta} 
A_{\alpha} ({\bf q}-{\bf q}')A_{\beta} ({\bf q}') \hat  \gamma_{\alpha \beta}(-{\bf q};2),
\label{eqB1}
 \end{eqnarray} 
and
\begin{eqnarray}
A_\mu ({\bf r}, t) = 
\left\{ \begin{array}{ll}
A_\alpha^{\rm ext} ({\bf r}, t), & \hspace{3mm} \mu = \alpha = 1, 2, 3 \\
& \\
cV^{\rm ext} ({\bf r}, t), &  \hspace{3mm} \mu = 0
\end{array} \right. .
\label{eqB2}
\end{eqnarray}
	The density operator in the second-order term, $\gamma_{\alpha \beta}(-{\bf q};2)$, 
is the bare diamagnetic density operator, and the $\gamma_{\alpha \beta}^{LL'} ({\bf k},{\bf k}_+;2)$ 
are the corresponding vertex functions.
	As pointed out in Sec.~IV, local charge conservation (\ref{eqA6}) follows as a consequence 
of gauge invariance of (\ref{eqB1}).

The result is
\begin{eqnarray}
&& \hspace{-5mm} 
J_\mu(q) = -\frac{1}{c} \sum_\nu \bigg(\pi_{\mu \nu}(q)  
+ \frac{e^2 n_{\mu \nu}^{\rm tot} ({\bf q})}{m}\bigg) A^{\rm tot}_\nu (q),
\label{eqB3}
\end{eqnarray}
with $n^{\rm tot}_{0\nu}=n^{\rm tot}_{\nu 0}= 0$ and
\begin{eqnarray}
&& \hspace{-5mm} n^{\rm tot}_{\alpha \beta} ({\bf q})
= \frac{1}{V} \sum_{L{\bf k} \sigma} \gamma_{\alpha \beta}^{LL} ({\bf k},{\bf k}_+ ;2) n_L({\bf k}).
\label{eqB4}
\end{eqnarray}
	The comparison of Eq.~(\ref{eqB4}) with Eqs.~(\ref{eq33}), (\ref{eq35}), and (\ref{eq36}) 
leads to the relation known as the effective mass theorem.
	For example, for the contribution to 
$\gamma_{\alpha \beta}^{LL} ({\bf k},{\bf k}_+;2) \approx \gamma_{\alpha \beta}^{LL} ({\bf k};2)$ 
which are diagonal in the polarization index $\alpha$, we obtain
\cite{Kupcic07,Kupcic13}
\begin{eqnarray}
&& \hspace{-10mm}
\gamma^{LL}_{\alpha \alpha}({\bf k};2) = \gamma^{LL}_{\alpha \alpha}({\bf k}) 
+ \frac{m}{e^2} \sum_{L'(\neq L)} \frac{2 J_\alpha^{L L'}({\bf k}) J_\alpha^{L'L}({\bf k})}{
\varepsilon_{L'L}({\bf k},{\bf k})},
\label{eqB5}
 \end{eqnarray} 
with  $\gamma_{\alpha \alpha}^{LL} ({\bf k}) = 
(m/\hbar^2) \partial^2 \varepsilon_L ({\bf k}) /\partial k_\alpha^2$ again.

The four-divergence of $J_\mu(q)$ (i.e., the charge continuity equation) reads as 
\begin{eqnarray}
&& \hspace{-5mm} 
\sum_\mu q_\mu J_\mu (q) = -\frac{1}{c} \sum_\nu A_\nu^{\rm tot} (q)
\sum_\mu q_\mu \bigg(\pi_{\mu \nu}(q)  + \frac{e^2 n_{\mu \nu}^{\rm tot} ({\bf q})}{m}\bigg) 
\nonumber \\
&& \hspace{15mm}
= 0.
\label{eqB6}
\end{eqnarray}
	Evidently this equation is fulfilled if, and only if Eq.~(\ref{eq39}) is satisfied.

An important feature of Eq.~(\ref{eqB3}) is that we can always choose the $V^{\rm tot} ({\bf r}, t)=0$ gauge, and write 
\cite{Schrieffer64}
\begin{eqnarray}
&& \hspace{-10mm} 
J_\alpha ({\bf q},\omega) = \sum_\beta \frac{{\it i}}{\omega} 
\big[\pi_{\alpha \beta}({\bf q},\omega) -\pi_{\alpha \beta}({\bf q}) \big] E_\beta ({\bf q},\omega)
\nonumber \\
&& \hspace{5mm}
-\frac{1}{c} \sum_\beta \bigg( \frac{e^2 n_{\alpha \beta}^{\rm tot}({\bf q})}{m} -
\pi_{\alpha \beta}({\bf q}) \bigg) A_\beta ({\bf q},\omega).
\label{eqB7}
\end{eqnarray}
	The first term is the paramagnetic contribution to the induced current originating from 
normal electrons and the second one is the diamagnetic contribution of superconducting electrons
(if the system under consideration is in the ordered superconducting state).

\subsection{Vertex functions}
In the electronic models described by the exactly solvable bare Hamiltonian
\begin{equation}
H_0^{\rm el} = \sum_{ll'} \sum_{{\bf k} \sigma} H^{ll'}_0 ({\bf k}) 
c^\dagger_{l {\bf k} \sigma} c_{l' {\bf k} \sigma}
=  \sum_{L{\bf k} \sigma} \varepsilon_L ({\bf k}) c^\dagger_{L {\bf k} \sigma} c_{L {\bf k} \sigma}
\label{eqB8}
\end{equation}
the vertex functions in Eq.~(\ref{eqB1}) are given by the general expressions
\cite{Kupcic00,Kupcic07}
\begin{eqnarray}
&& \hspace{-5mm}
q^{LL'}({\bf k},{\bf k}_+) = \sum_{ll'} q^{ll'}({\bf k},{\bf k}_+) U_{\bf k} (l,L) U_{{\bf k}+{\bf q}}^* (l',L'),
\nonumber \\
&& \hspace{-5mm}
J^{LL'}_{\alpha}({\bf k},{\bf k}_+) = \sum_{ll'} \frac{e}{\hbar} 
\frac{\partial  H_0^{ll'} ({\bf k})}{\partial k_{\alpha}} \, U_{\bf k} (l,L) U_{{\bf k}+{\bf q}}^* (l',L'),
\nonumber \\
&& \hspace{-5mm}
\gamma^{LL'}_{\alpha \beta}({\bf k},{\bf k}_+;2) = \sum_{ll'} \frac{m}{\hbar^2} 
\frac{\partial^2  H_0^{ll'} ({\bf k})}{\partial k_{\alpha} \partial k_\beta} \,
U_{\bf k} (l,L) U_{{\bf k}+{\bf q}}^* (l',L'),
\nonumber \\
\label{eqB9}
 \end{eqnarray} 
with $\rho^{ll'}({\bf k},{\bf k}_+) \approx \delta_{l,l'}$.
	Here, the $U_{\bf k} (l,L)$ are the transformation matrix elements in 
$c^\dagger_{L {\bf k} \sigma} = \sum_l U_{\bf k} (L,l) c^\dagger_{l {\bf k} \sigma}$.
	The sum $\sum_l$ runs over all orbitals in the unit cell which participate 
in building up the valence bands.

\section{Vertex functions in graphene}
As mentioned in the main text, in the case in which the overlap parameter $s$ is set equal to zero,
the relevant matrix elements $H^{ll'}_0 ({\bf k})$ in graphene are
$H^{ll}_0 ({\bf k}) = \varepsilon_{p_z}=0$ and $H^{BA}_0 ({\bf k}) = t({\bf k})$.
	Thus the transformation matrix between the delocalized orbital states
$|l {\bf k} \sigma \rangle = c^\dagger_{l {\bf k} \sigma} |0\rangle$  ($l = A, B$) and the Bloch states
$|s {\bf k} \sigma \rangle = c^\dagger_{s {\bf k} \sigma} |0\rangle$ ($s = \pi^*, \pi$) are
\cite{Castro09}
\begin{equation}
\left( \begin{array}{cc} U_{{\bf k}} (A,\pi^*) & U_{{\bf k}} (A,\pi) \\
 U_{{\bf k}} (B,\pi^*) & U_{{\bf k}} (B,\pi) \end{array} \right) 
= \frac{1}{\sqrt{2}} \left( \begin{array}{cc} 1 & 1 \\
 {\rm e}^{-{\rm i} \theta_{\bf k}} & - {\rm e}^{-{\rm i} \theta_{\bf k}} \end{array} \right).
\label{eqC1}
\end{equation}
	The auxiliary phase $\theta_{\bf k}$ is defined by
$\tan \theta_{\bf k} = t_i ({\bf k})/t_r ({\bf k})$, with $t_r ({\bf k})$ and $t_i ({\bf k})$
being the real and the imaginary part of $t ({\bf k})$.

By substituting this expression into Eqs.~(\ref{eqB9}), we obtain the monopole-charge vertex functions \cite{Hwang07}
\begin{eqnarray}
&& \hspace{-10mm}
q^{ss'}({\bf k},{\bf k}') 
= \frac{1}{2} \big(1 + s s' e^{i(\theta_{{\bf k}'}-\theta_{\bf k} )} \big).
\label{eqC2}
\end{eqnarray} 
	Similarly, it is not hard to verify that the monopole-charge vertices and the current vertices satisfy 
the general relation (\ref{eqA3}), resulting in 
\begin{eqnarray}
&& \hspace{-10mm}
q_\alpha \hbar J_\alpha^{ss'} ({\bf k},{\bf k}') = 
(s'|t({\bf k}')|-s|t({\bf k})|) e q^{ss'} ({\bf k}, {\bf k}')
\label{eqC3}
\end{eqnarray}
(${\bf q} = q_\alpha \hat e_\alpha$ and  $q_\alpha =k'_\alpha-k_\alpha$).

For long wavelengths, we obtain 
\begin{eqnarray}
&& \hspace{-10mm}
q^{ss'}({\bf k},{\bf k}_+) 
\approx \frac{1}{2}\bigg[1 + ss'\bigg(1 + {\it i}q_\alpha \frac{\partial \theta_{{\bf k}}}{\partial k_\alpha} \bigg) \bigg],
\label{eqC4}
\end{eqnarray} 
and \cite{Ziegler06,Lewkowicz09}
\begin{eqnarray}
 && \hspace{-10mm}
J^{ss'}_{\alpha} ({\bf k},{\bf k}_+) \approx J^{ss'}_{\alpha} ({\bf k}) 
\nonumber \\
&& \hspace{8mm}
= s \frac{e}{\hbar 2 |t({\bf k})|} 
\bigg[t^*({\bf k}) \frac{\partial t ({\bf k})}{\partial k_\alpha} 
+ s s' t({\bf k}) \frac{\partial t^* ({\bf k})}{\partial k_\alpha} \bigg].
\nonumber \\
\label{eqC5}
\end{eqnarray} 
	The latter expression can also be written in the form
\begin{eqnarray}
 && \hspace{-10mm}
J^{ss}_{\alpha} ({\bf k}) = s \frac{e}{\hbar} \frac{\partial |t ({\bf k})|}{\partial k_\alpha}
= e v^{s}_{\alpha} ({\bf k}) = \frac{e}{\hbar} \frac{\partial E_s ({\bf k})}{\partial k_\alpha},
\nonumber \\
&& \hspace{-10mm}
J^{s \underline{s}}_{\alpha} ({\bf k}) =  
s \frac{i e \, |t ({\bf k})|}{\hbar} \frac{\partial \theta_{\bf k}}{\partial k_\alpha}.
\label{eqC6}
 \end{eqnarray} 
	Alternatively, we can write
\begin{eqnarray}
 && \hspace{-10mm}
J^{ss}_{\alpha} ({\bf k}) = s \frac{e t a}{2 \hbar} \frac{1}{|\tilde t ({\bf k})|} j^{\rm intra}_{\alpha} ({\bf k}),
\nonumber \\
&& \hspace{-10mm}
J^{s \underline{s}}_{\alpha} ({\bf k}) =  s i \frac{e t a}{2 \hbar} \frac{1}{|\tilde t ({\bf k})|} j^{\rm inter}_{\alpha} ({\bf k}),
\label{eqC7}
 \end{eqnarray} 
where 
\cite{PelcUP}
\begin{eqnarray}
&& \hspace{-5mm}
j^{\rm intra}_{x} ({\bf k}) = -2 \bigg( \sin k_xa + \sin\frac{k_xa}{2} \cos \frac{\sqrt{3} k_ya}{2}\bigg),
\nonumber \\
&& \hspace{-5mm}
j^{\rm intra}_{y} ({\bf k}) = - 2\sqrt{3}\cos\frac{k_xa}{2} \sin\frac{\sqrt{3} k_ya}{2},
\nonumber \\
&& \hspace{-5mm}
j^{\rm inter}_{x} ({\bf k}) = 2\sin\frac{k_xa}{2} \sin\frac{\sqrt{3} k_ya}{2},
\nonumber \\
&& \hspace{-5mm}
j^{\rm inter}_{y} ({\bf k}) = 
-\frac{2\sqrt{3}}{3} \bigg( \cos k_xa - \cos\frac{k_xa}{2} \cos \frac{\sqrt{3} k_ya}{2} \bigg),
\nonumber \\
\label{eqC8}
 \end{eqnarray} 
and $|\tilde t ({\bf k})| = |t ({\bf k})/t|$.
	Finally, the elements of the reciprocal effective mass tensor can be written in the form
\begin{eqnarray}
&& \hspace{-5mm}
\gamma^{ss}_{\alpha \beta} ({\bf k}) = s \frac{m}{m_{xx}} \frac{1}{a} \frac{\partial }{\partial k_\beta}
j^{\rm intra}_{\alpha} ({\bf k}),
\label{eqC9}
 \end{eqnarray} 
where $m_{xx} = (2 \hbar^2/t a^2)$.

In the Dirac cone approximation in the vicinity of the K point ($\tilde {\bf k} ={\bf k} -{\bf k}_K$), 
these expressions reduce to 
\begin{eqnarray}
&& \hspace{-10mm}
J^{ss}_{\alpha} ({\bf k}) = s e v_{\rm F} \tilde k_\alpha / \tilde k, 
\nonumber \\
&& \hspace{-10mm}
J^{s \underline{s}}_{x} ({\bf k}) = si e v_{\rm F} \tilde k_y / \tilde k \equiv {\it i} J^{ss}_{y} ({\bf k}), 
\nonumber \\
&&
\hspace{-10mm}
J^{s \underline{s}}_{y} ({\bf k}) = -s i e v_{\rm F} \tilde k_x / \tilde k \equiv -{\it i} J^{ss}_{x} ({\bf k}), 
\label{eqC10}
\end{eqnarray}
and
\begin{eqnarray}
&& \hspace{-5mm}
\gamma^{ss}_{\alpha \alpha} ({\bf k}) = s \frac{\sqrt{3}m}{m_{xx}} 
\bigg( \frac{1}{\tilde k a} - \frac{(\tilde k_\alpha a)^2}{(\tilde k a)^3}  \bigg).
\label{eqC11}
 \end{eqnarray} 

\subsection{Electron-phonon vertex functions}
The coupling between conduction electrons and in-plane optical phonons
in Eq.~(\ref{eq12}) is given by $g_\nu \equiv g = - \partial t_j /\partial r_j$ and
\cite{Ando06,Peres08,Kupcic12}
\begin{eqnarray}
q_{\nu }^{ss'} ({\bf k}_+,{\bf k}) &=& \sum_{l \neq l'} 
U_{{\bf k+q}} (l,s) U^*_{{\bf k}} (l',s') q^{ll'}_\nu  ({\bf k}_+,{\bf k}), 
\nonumber \\
q^{AB}_\nu  ({\bf k}_+,{\bf k}) &=& 
-\sum_{j=1}^3 {\bf r}_{j0} \cdot {\bf e}_{\bf q}^\nu \left( 1 
+ {\rm e}^{{\rm i} {\bf q} \cdot ({\bf r}_3 - {\bf r}_j )} \right) {\rm e}^{-{\rm i} {\bf k} \cdot {\bf r}_j}.
\nonumber \\
\label{eqC12}
\end{eqnarray}
	For ${\bf q} = q_\alpha \hat e_\alpha$, $q_\alpha \approx 0$ and $\beta \in \{ \alpha, \underline{\alpha} \}$, 
the direct calculation gives $q_{\beta }^{ss'} ({\bf k}_+,{\bf k}) \approx q_{\beta }^{ss'} ({\bf k})$ and 
\begin{eqnarray}
 && \hspace{-10mm}
q_{\beta }^{ss} ({\bf k}) \approx \frac{3 {\it i}}{e v_{\rm F}} J^{s\underline{s}}_{\beta} ({\bf k}),
\hspace{5mm}
q_{\beta }^{s\underline{s}} ({\bf k}) \approx \frac{3 {\it i}}{e v_{\rm F}} J^{ss}_{\beta} ({\bf k}).
\label{eqC13}
 \end{eqnarray} 
	In the Dirac cone approximation, this leads to \cite{Ando06,Castro07}
\begin{eqnarray}
 && \hspace{-10mm}
q_{x}^{ss'} ({\bf k}) \approx -\frac{3}{e v_{\rm F}} J^{ss'}_{y} ({\bf k}),
\hspace{5mm}
q_{y}^{ss'} ({\bf k}) \approx \frac{3}{e v_{\rm F}} J^{ss'}_{x} ({\bf k}).
\label{eqC14}
 \end{eqnarray} 
%

% \newpage  

\end{document}